\documentclass[reprint,aps,prl,nobibnotes,superscriptaddress]{revtex4-1}
\usepackage[T1]{fontenc}
\usepackage[latin9]{inputenc}
\usepackage{float}
\usepackage{bm}
\usepackage{amsmath}
\usepackage{printlen}
\usepackage{amssymb}
\usepackage{graphicx, xcolor}
\usepackage{babel}
\usepackage{hyperref,xr}
\begin{document}
\title{Light propagation and magnon-photon coupling in optically dispersive magnetic media}

\author{V. A. S. V. Bittencourt}
\affiliation{Max Planck Institute for the Science of Light, 91058 Erlangen, Germany}

\author{I. Liberal}
\affiliation{Electrical and Electronic Engineering Department, Institute of Smart Cities (ISC), Universidad P\'{u}blica de Navarra (UPNA), 31006 Pamplona, Spain}

\author{S. {Viola Kusminskiy}}
\affiliation{Max Planck Institute for the Science of Light, 91058 Erlangen, Germany}
\affiliation{Department of Physics, University Erlangen-N{\"u}rnberg, 91058 Erlangen, Germany} 

\begin{abstract}
Achieving strong coupling between light and matter excitations in hybrid systems is a benchmark for the implementation of quantum technologies. We recently proposed [arXiv:2110.02984] that strong single-particle coupling between magnons and light can be realized in a magnetized epsilon-near-zero (ENZ) medium, in which magneto-optical effects are enhanced. 
Here we present a detailed derivation of the magnon-photon coupling Hamiltonian in dispersive media both for degenerate and non-degenerate optical modes, and show the enhancement of the coupling near the ENZ frequency. Moreover, we show that the coupling of magnons to plane-wave non-degenerate Voigt modes vanishes at specific frequencies due to polarization selection rules tuned by dispersion. Finally, we present specific results using a Lorentz dispersion model. Our results pave the way for the design of dispersive optomagnonic systems, providing a general theoretical framework for describing  engineering ENZ-based optomagnonic systems.
\end{abstract}

\maketitle

\section{Introduction}

Achieving strong coupling between light and matter has been a benchmark for the development of quantum technologies \cite{Diaz_2019_Ultrastrong,Frisk_2019_ultrastrong}, specially for the design of quantum hybrid systems \cite{Kurizki_2015_quantum_technologies}. The strong coupling regime is characterized by a coupling rate that is larger than the typical decay rates of the constituents, allowing the exchange of quanta between light and matter excitations in a timescale faster than the decays. This in turn allows the implementation of quantum protocols such as state transfer \cite{Diaz_2019_Ultrastrong}. Systems in which the strong coupling regime has been successfully achieved include the coupling of superconducting qubits to microwaves \cite{Wallraff_2004_ultrastrong,Forn_2010_observation_Bloch,Yoshihara_2017_superconducting}, semiconductors to ${\rm THz}$ light \cite{Kusch_2021_strong_light,Peraca_2020_ultrastrong} and the magnetization of dielectrics to microwave cavity photons \cite{Huebl_2013_high_coop,Tabuchi_2014_Hybridizing_Ferromagnetic,Zhang_2014_Strongly_Coupled,Bourhill_2016_Ultrahigh_cooperativity}. The latter type of systems, called cavity magnonic systems, are particularly attractive since the magnetization can also couple to optical photons \cite{Rameshti_2021_Cavity_Magnonics}, opening the prospect of a magnonic-based microwave-to-opical quantum transducer\cite{Hisatomi_2016_Biderectional_conversion,Lachance_Quirion_2019_Hybrid_Quantum,Zhu_2020_Waveguide_cavity,Lambert_2020_Coherent_Conversion}.

While the strong coupling regime between the quantized magnetic excitations (magnons) and microwave photons of a cavity has been now achieved in different experimental setups \cite{Tabuchi_2014_Hybridizing_Ferromagnetic,Zhang_2014_Strongly_Coupled,Bourhill_2016_Ultrahigh_cooperativity,Huebl_2013_high_coop}, the coupling to optical photons is still very weak \cite{Haigh_2020_subpicoliter,Zhu_2020_Waveguide_cavity,Zhu_2021_inverse}, limiting the applications of such systems. The weakness of the optomagnonic coupling is partly a consequence of the weakness of the magneto-optical (MO) effects, such as the Faraday and the Cotton-Mouton effects, responsible for the coupling between light and magnetization. This roadblock can be partially bypassed by considering an optomagnonical cavity \cite{Kusminskiy_2016_Coupled,Haigh_2016_Triple_Resonant, Zhang_2016_Optomagnonic_whispering,Liu_2016_Optomagnonics}: a setup in which the light is trapped inside the magnetic dielectric by total internal reflection, yielding a cavity enhancement of the coupling. Efforts are being made to optimize such systems \cite{Graf_2018_Cavity_optomagnonics, Graf_2021_Design_optomagnonic}, with significant experimental and theoretical advancements in the recent years. Nevertheless, the coupling in state-of-the-art systems ($\sim50$ Hz \cite{Haigh_2020_subpicoliter}) is not strong enough to surpass the decay rates of photons ($\sim1$GHz) and magnons ($\sim1$ MHz). Theoretical proposals for optomagnonic crystals predict a stronger coupling rate ($\sim 10$ kHz) \cite{Graf_2021_Design_optomagnonic}, which is still several orders of magnitude smaller than the optical decay rate in such systems ($\sim0.1$ THz).

Here we explore an alternative route for achieving single particle strong coupling between magnons and photons on an epsilon-near-zero (ENZ) medium, as proposed in Ref.~\cite{Bittencourt_2021_ENZ_Lett}. ENZ media exhibit an almost vanishing permittivity at a frequency called the epsilon-near-zero frequency \cite{liberal_near_zero_2017}. This behavior is a consequence of dispersion, either due to the bulk material (e.g., the usual Lorentz dispersion of dielectrics) or due to the structure of the medium (e.g., a metallic wave-guide operating at a cutoff frequency). In such media, non-linear effects and secondary responses of the media to optics are enhanced, making ENZ, and more generally speaking near-zero-index media, interesting platforms for matter-light interactions and hybrid systems, for instance, allowing the realization of strong plasmon-phonon coupling \cite{Yoo_2021_ultrastrong_plasmon}. In particular, MO effects in ENZ media have been predicted to yield perfect optical isolation based on the intrinsic non-reciprocal behavior of magnetic systems \cite{davoyan_optical_isolation_2013,davoyan_nonreciprocal_emission_2019}, unidirectional propagation of photonic states at interfaces between dielectrics and metals \cite{davoyan_theory_of_2013}, and giant transverse MO Kerr effect \cite{Almpanis_controllingTMOKE_2020,giron_giant_enhancement_2017}. Such enhancement of MO effects points to the possibility of an enhanced magnon-photon coupling at the single-particle level in ENZ media, but its description requires a theoretical framework that incorporates dispersion, in particular a quantization procedure to describe photons in magnetic dielectrics, magnons, and their coupling.

In this manuscript, we present a detailed derivation of the magnon-photon coupling in dispersive magnetic dielectrics. We extend the results presented in Ref. \cite{Bittencourt_2021_ENZ_Lett} for the magnon-photon coupling to include non-degenerate optical modes and we treat both the Faraday and Voigt configurations (light propagating parallel or perpendicular to the magnetization, respectively). We derive the general features of the coupling, including the coupling enhancement and the linear relation between coupling and the ENZ frequency reported in  Ref. \cite{Bittencourt_2021_ENZ_Lett}. For non-degenerate modes, we show that the coupling constant can vanish at specific frequencies due to \textquotedbl dispersion-tuned selection rules\textquotedbl{} that allow only the up or down conversion of photons with specific polarizations. Such features are independent of the dispersion model and can be applied both to ENZ materials or to structured media. Finally, we specialize our results to a dispersion model inspired by the dispersion in bulk paramagnetic and ferromagnetic dielectrics. We study the propagation regimes of plane-waves in such framework, recovering the isolation features for waves propagating in the Faraday configuration at the ENZ regime reported in the literature \cite{davoyan_optical_isolation_2013}, as well as in the Voigt configuration, and evaluate the magnon-photon coupling, showing the coupling enhancement at ENZ and the dispersion selection rules for non-degenerate modes. Similar dispersive models can be found in magnetized plasmas \cite{Shen_analogof_2019} and metal-gyroelectric dielectric structures \cite{Tsakmakidis_2017_breaking_lorentz}.

Our formalism can be applied to any magnetic dispersive medium, bulk or structured, provided that the intrinsic losses are small. Different from previous studies of MO effects in ENZ media, which are focused on entirely classical phenomena \cite{davoyan_optical_isolation_2013,davoyan_nonreciprocal_emission_2019,davoyan_theory_of_2013,giron_giant_enhancement_2017,Almpanis_controllingTMOKE_2020}, our quantized framework is aimed to describe the coupling between single magnons and single photons in order to explore the quantum regime. Our results push forward the design of optomagnonic systems and further motivate the use of ENZ media as platforms for quantum technologies and light-matter interactions. Dispersion can also lead to non-Markovian phenomena \cite{Genes_2019_cavity_quantum,Liberal_2021_nonperturbative,Gonzales_2017_Quantum_Emitters} which adds novel aspects to optomagnonics.

This paper is structured as follows. In Section \ref{sec:Quantization-of-optomagnonics} we first obtain the energy densities including dispersion in Section \ref{subsec:Energy-density}. Then in Section \ref{subsec:Quantization-of-the} we quantize the electromagnetic field modes following Ref. \cite{Milonni_1995_Field_Quantization} and extend the formalism to obtain the optomagnonic coupling Hamiltonian. In Section \ref{subsec:Plane-wave-like-modes:} we derive the relevant optical polarization for plane waves propagating in magnetized media and for modes of a magneto-optical Fabry-P\'{e}rot cavity. With those modes we evaluate the optomagnonic Hamiltonian for degenerate and non-degenerate modes in Section \ref{subsec:Optomagnonic-coupling-Hamiltonia}, where we also derive some general features of the coupling, such as its behavior at the ENZ frequency and the requirements for achieving strong single magnon-photon coupling. Finally, in Section \ref{sec:Lorentz-dispersion-for} we present a dispersion model for paramagnetic and ferromagnetic dielectrics, which we use to specify our results for the optomagnonic coupling in Section \ref{subsec:Optomagnonic-coupling-for} for both degenerate and non-degenerate modes. In Section \ref{sec:Conclusions} we present the conclusions and outlook.

\section{Magnon-photon coupling Hamiltonian including dispersion\label{sec:Quantization-of-optomagnonics}}

In what follows, we present a detailed derivation of the Hamiltonian presented in \cite{Bittencourt_2021_ENZ_Lett} and  generalize the framework for describing coupling between magnons and non-degenerate optical  modes. 

\subsection{Energy density\label{subsec:Energy-density}}

The starting point is the energy density in a dispersive medium, obtained following \cite{Landau_electrodynamics_2009,Stancil_spin_2009}. From the instantaneous Poynting theorem, the energy density is given by
\begin{equation}
w(t)=\int_{-\infty}^{t}dt^{\prime}\left[\bm{H}\cdot\dot{\bm{B}}+\bm{E}\cdot\dot{\bm{D}}\right],
\end{equation}
where we $\bm{E}$ and $\bm{D}$ are the electric and displacement fields and $\bm{H}$ and $\bm{B}$ are the magnetic and induction fields. We split $w(t)$ into an electric part $w_{e}(t)=\int_{-\infty}^{t}dt^{\prime}\left[\bm{E}\cdot\dot{\bm{D}}\right]$ and a magnetic part $w_{m}(t)=\int_{-\infty}^{t}dt^{\prime}\left[\bm{H}\cdot\dot{\bm{B}}\right]$. The energy density here only takes into account the optical fields and the magnetization. To describe the magnetization's dynamics, one has to include magnetization-dependent effective fields describing, for example, exchange interactions and crystalline anisotropy, plus any other external field, such as an external bias magnetic field \cite{Stancil_spin_2009,White_QuantumTheoryMag_2007}. The magnetization's dynamics is then governed by the so-called Landau-Lifshitz equation, which is, in general, a non-linear equation for the magnetization. For small magnetic excitations, that is, for small deviations from the magnetic ground state, the magnetic excitations behave as harmonic oscillators, which when quantized describe bosonic quasi-particles called magnons \cite{Stancil_spin_2009}. The part of the energy density corresponding to the magnetization's dynamics yields in this limit a harmonic oscillator Hamiltonian.

Turning back our attention to the energy density of the electromagnetic field, let us first consider a monochromatic field with slowly varying amplitude:
\begin{equation}
\begin{aligned}
\bm{E}(\bm{r},t)&={\rm Re}\left\{ \bm{E}^{(+)}(\bm{r},t)\right\} ={\rm Re}\left\{ \bm{E}^{(+)}(\bm{r})e^{-is_{c}t}\right\}\\ &=\left(\bm{E}^{(+)}(\bm{r})e^{-is_{c}t}+\bm{E}^{(-)}(\bm{r})e^{is_{c}^{*}t}\right)/2,
\end{aligned}
\label{eq:ElectricField}
\end{equation}
where $s_{c}=\omega_{c}+i\alpha_{c}$ and $\alpha_{c}\ll\omega_{c}$. The displacement field is given by $\bm{D}(\bm{r},t)={\rm Re}\left\{ \varepsilon_{0}\overleftrightarrow{\varepsilon}[s_{c},\bm{\mathcal{M}}]\cdot\bm{E}^{(+)}(\bm{r})e^{-is_{c}t}\right\} $, where $\overleftrightarrow{\varepsilon}[\omega,\bm{\mathcal{M}}]$ is the permittivity tensor that depends on the magnetization $\bm{\mathcal{M}}$. The permittivity tensor satisfies the following properties: (i) from its definition in terms of a response function $\overleftrightarrow{\varepsilon}^{*}[\omega,\bm{\mathcal{M}}]=\overleftrightarrow{\varepsilon}[-\omega,\bm{\mathcal{M}}]$; (ii) we assume frequency ranges in which absorption is small, such that $\overleftrightarrow{\varepsilon}^{\dagger}[\omega,\bm{\mathcal{M}}]=\overleftrightarrow{\varepsilon}[\omega,\bm{\mathcal{M}}]$; and (iii) the Onsager reciprocity relations imply that ${\rm Re}\left\{ \varepsilon_{ij}[\omega,\bm{\mathcal{M}}]\right\} ={\rm Re}\left\{ \varepsilon_{ji}[\omega,-\bm{\mathcal{M}}]\right\} $ and ${\rm Im}\left\{ \varepsilon_{ij}[\omega,\bm{\mathcal{M}}]\right\} ={\rm Im}\left\{ \varepsilon_{ji}[\omega,-\bm{\mathcal{M}}]\right\} $. To first order in the magnetization this implies the generic form
\begin{equation}
\overleftrightarrow{\varepsilon}[\omega,\bm{\mathcal{M}}]=\left[\begin{array}{ccc}
\varepsilon[\omega] & i\mathcal{F}[\omega]\mathcal{M}_{z} & -i\mathcal{F}[\omega]\mathcal{M}_{y}\\
-i\mathcal{F}[\omega]\mathcal{M}_{z} & \varepsilon[\omega] & i\mathcal{F}[\omega]\mathcal{M}_{x}\\
i\mathcal{F}[\omega]\mathcal{M}_{y} & -i\mathcal{F}[\omega]\mathcal{M}_{x} & \varepsilon[\omega]
\end{array}\right].\label{eq:PermitividadeEletrica}
\end{equation}
We will consider small fluctuations around the saturation magnetization, such that $\mathcal{M}_{z} \sim \mathcal{M}_S$ and $\mathcal{M}_{x,y} \ll \mathcal{M}_{S}$, where $\mathcal{M}_{S}$ is the saturation magnetization, and we assume that the magnetization and its fluctuations are uniform. Eq.~\eqref{eq:PermitividadeEletrica} takes into account only magnetic (circular) birefringence, described by the off-diagonal elements of $\overleftrightarrow{\varepsilon}$, generated by the Faraday effect. The Cotton-Mouton effect, which generates both linear and circular magnetic birefrigence, can be taken into account by adding to Eq.~(\ref{eq:PermitividadeEletrica}) second order terms in the magnetization, which we do not consider in this work. For now, we do not specify $\varepsilon[\omega]$ and $\mathcal{F}[\omega]$. In Section \ref{sec:Lorentz-dispersion-for} we will illustrate our results for a specific model of dispersion in which $\varepsilon[\omega]$ is described by the Lorentz model and $\mathcal{F}[\omega]$ describes the dispersion of the Faraday effect as obtained in paramagnetic and ferromagnetic dielectrics.

The electric part of the energy density is given by
\begin{widetext}
\begin{equation}
w_{e}(t)=\frac{1}{4}\int_{-\infty}^{t}dt^{\prime}(-is_{c})\left(\bm{E}^{(+)}(\bm{r})\cdot\bm{D}^{(+)}(\bm{r})e^{-2i(\omega_{c}-i\alpha_{c})t^{\prime}}+\bm{E}^{(-)}(\bm{r})\cdot\bm{D}^{(+)}(\bm{r})e^{2\alpha_{c}t^{\prime}}+{\rm c.c.}\right).\label{eq:FullEDenE}
\end{equation}
Taking the average over a period of the monochromatic wave and discarding the terms $\propto e^{\pm2i\omega_{c}t^{\prime}}$ we obtain 
\begin{equation}
u_{e}=\langle w_{e}(t)\rangle=\frac{\varepsilon_{0}}{4}\int_{-\infty}^{t}dt^{\prime}\left(\bm{E}^{(-)}(\bm{r})\cdot\left(-is_{c}\overleftrightarrow{\varepsilon}[s_{c},\bm{\mathcal{M}}]\right)\cdot\bm{E}^{(+)}(\bm{r})e^{2\alpha_{c}t^{\prime}}+{\rm c.c.}\right)\,\label{eq:RWAEnergy}
\end{equation}
corresponding to a rotating wave approximation (RWA). We consider small dissipation, $\alpha_{c}\ll\omega_{c}$, and expand the permittivity tensor up to linear order in $\alpha_{c}$
\begin{equation}
-is_{c}\overleftrightarrow{\varepsilon}[s_{c},\bm{\mathcal{M}}]\sim-i\omega_{c}\overleftrightarrow{\varepsilon}[\omega_{c},\bm{\mathcal{M}}]+\alpha_c \partial_{\omega_{c}}\left(\omega_{c}\overleftrightarrow{\varepsilon}[\omega_{c},\bm{\mathcal{M}}]\right),
\end{equation}
and thus
\begin{equation}
u_{e}\approx\frac{\varepsilon_{0}}{2}\int_{-\infty}^{t}dt^{\prime}e^{2\alpha_{c}t^{\prime}} \left( \alpha_{c}\bm{E}^{(-)}(\bm{r})\cdot\partial_{\omega_{c}}\left(\omega_{c}\overleftrightarrow{\varepsilon}[\omega_{c},\bm{\mathcal{M}}]\right)\cdot\bm{E}^{(+)}(\bm{r}) \right).
\end{equation}
After integrating over time and taking the limit $\alpha_{c}\rightarrow0$ we obtain
\begin{equation}
u_{e}=\frac{\varepsilon_{0}}{4}\bm{E}^{(-)}(\bm{r},t)\cdot\partial_{\omega_{c}}\left(\omega_{c}\overleftrightarrow{\varepsilon}[\omega_{c},\bm{\mathcal{M}}]\right)\cdot\bm{E}^{(+)}(\bm{r},t),
\end{equation}
where we have used the definition of the monochromatic field Eq.~\eqref{eq:ElectricField}.

For the magnetic part of the energy density, we recall that in a magnetic material $\bm{B}=\mu_{0}\left(\bm{H}+\bm{\mathcal{M}}\right)$, where $\bm{\mathcal{M}}$ is the magnetization. For a monochromatic field $\bm{B}={\rm Re}\left\{ \bm{B}^{(+)}(\bm{r})e^{-is_{c}t}\right\} $, while for the magnetization we do not assume a linear response to the magnetic field. Instead, we are interested on the coupling between light and magnetic excitations, the latter described by
\begin{equation}
\bm{\mathcal{M}}=\bm{\mathcal{M}}_{s}+{\rm Re}\left\{ \delta\bm{\mathcal{M}}_{+}(\bm{r})e^{-is_{m}t}\right\} \label{eq:MagnExcit}
\end{equation}
where $s_{m}=\omega_{m}+i\alpha_{m}$, and $\omega_{m}$ is the frequency of the relevant magnon mode. This yields
\begin{equation}
\begin{aligned}
w_{m}(t) & =\frac{1}{4\mu_{0}}\int_{-\infty}^{t}dt^{\prime}\left(-is_{o}\bm{B}^{(-)}(\bm{r})\cdot\bm{B}^{(+)}(\bm{r})e^{2\alpha_{c}t^{\prime}}-is_{o}\bm{B}^{(+)}(\bm{r})\cdot\bm{B}^{(+)}(\bm{r})e^{-2i(\omega_{c}-i\alpha_{c})t^{\prime}}+{\rm c.c.}\right)\\
 & \:-\frac{1}{4\mu_{0}}\int_{-\infty}^{t}dt^{\prime}\left(-is_{m}\bm{B}^{(-)}(\bm{r})\cdot\delta\bm{\mathcal{M}}_{+}(\bm{r})e^{-i(\omega_{m}-\omega_{c})t}e^{(\alpha_{c}+\alpha_{m})t}+{\rm c.c.}\right)\\
 & \,-\frac{1}{4\mu_{0}}\int_{-\infty}^{t}dt^{\prime}\left(-is_{m}\bm{B}^{(+)}(\bm{r})\cdot\delta\bm{\mathcal{M}}_{+}(\bm{r})e^{-i(\omega_{m}+\omega_{c})t}e^{(\alpha_{c}+\alpha_{m})t}+{\rm c.c.}\right).
\end{aligned}
\end{equation}
\end{widetext}
We consider the typical magnetic excitation frequency, which is in the GHz range. Since $\omega_{c}$ is at an optical frequency, there is a frequency mismatch $\omega_{c}\gg\omega_{m}$ and after we take the time average of $w_{m}(t)$ over one period of oscillations of the optical electromagnetic field, the terms on the second and third lines can be discarded in a RWA. We follow the same procedure that leads to $u_{e}$, and we get for the magnetic field contribution
\begin{equation}
u_{m}=\frac{1}{4\mu_{0}}\bm{B}^{(-)}(\bm{r},t)\cdot\bm{B}^{(+)}(\bm{r},t).
\end{equation}

The total energy density for a monochromatic wave at frequency $\omega_{c}$ thus reads
\begin{equation}
\begin{aligned}
u_{\omega_{c}} & =\frac{1}{4}\Big(\varepsilon_{0}\bm{E}^{(-)}(\bm{r},t)\cdot\partial_{\omega_{c}}\left(\omega_{c}\overleftrightarrow{\varepsilon}[\omega_{c},\bm{\mathcal{M}}]\right)\cdot\bm{E}^{(+)}(\bm{r},t) \\
&\quad \quad +\frac{1}{\mu_{0}}\bm{B}^{(-)}(\bm{r},t)\cdot\bm{B}^{(+)}(\bm{r},t)\Big).
\label{eq:EnergyDensityDispersion}
\end{aligned}
\end{equation}
This is the energy density we use as a starting point for our quantization procedure in the case of a monochromatic field that yields the magnon-photon coupling including dispersion. The definition of energy density in a dispersive medium was thoroughly discussed in \cite{Rosa_2010_electromagnetic_energy}, and we take it as a starting point for a phenomenological quantization of the optomagnonic coupling in the presence of dispersion, as done in \cite{Bittencourt_2021_ENZ_Lett}.

\subsubsection{Non-degenerate modes}

We can now adapt the formalism developed in the previous section to describe many non-degenerate optical modes that couple to the magnetic excitations. In this case, we assume that the total electromagnetic field in the volume of interest is a superposition of discrete frequency components $\bm{E}(\bm{r},t)=\sum_{c}{\rm Re}\left\{ \bm{E}_{c}^{(+)}(\bm{r},t)\right\} =\sum_{c}{\rm Re}\left\{ \bm{E}_{c}^{(+)}(\bm{r})e^{-is_{c}t}\right\} $. 

Following the same procedure of last section, we obtain for the magnetic part of the energy density
\begin{equation}
u_{m}=\frac{1}{4\mu_{0}}\sum_{c}\bm{B}_{c}^{(-)}(\bm{r},t)\cdot\bm{B}_{c}^{(+)}(\bm{r},t).
\end{equation}
For the electric part of the energy density we consider as before the displacement field obtained via the permittivity tensor. We are interested in effects due to magnetization's fluctuations around it's saturation value, described by the magnetization's fluctuations in Eq.~\eqref{eq:MagnExcit} included only up to first order in the permittivity, yielding
\begin{equation}
\begin{aligned}
\overleftrightarrow{\varepsilon}[\omega_{c},\bm{\mathcal{M}}] &=\overleftrightarrow{\varepsilon}_{S}[\omega_{c},\bm{\mathcal{M}}_{S}] \\
&+\left(\overleftrightarrow{\varepsilon}_{+}[\omega_{c},\delta\bm{\mathcal{M}}_{+}]e^{-is_{m}t}+ \rm{h.c.}\right).
\end{aligned}
\end{equation}
This corresponds to splitting the permittivity tensor in a static part and a fluctuating part, see for example \cite{Pantazopoulos_2017_Photomagnonic_nanocavities,Almpanis_2020_Spherical_optomagnonic}. We consider that the saturation magnetization is aligned along $\bm{e}_{z}$, such that $\delta\bm{\mathcal{M}}_{\pm}=(\mathcal{M}_{x} \mp i \mathcal{M}_{y})\bm{e}_{\pm}/\sqrt{2}$, with $\bm{e}_{\pm} = (\bm{e}_x \pm i \bm{e}_y)/\sqrt{2}$. We also define $\overleftrightarrow{\varepsilon}_{+}^{\dagger}[\omega_{c},\delta\bm{,\mathcal{M}}_{+}]=\overleftrightarrow{\varepsilon}_{-}[\omega_{c},\delta\bm{\mathcal{M}}_{-}]$. 

The electric energy density thus decomposes into two parts
\begin{equation}
w_{e}(t)=w_{e,S}(t)+w_{e,\delta}(t),
\end{equation}
where
\begin{widetext}
\begin{equation}
\begin{aligned}
w_{e,S}(t) & =\frac{\varepsilon_{0}}{4}\int_{-\infty}^{t}dt^{\prime}\sum_{c,c^{\prime}}\Big[(-is_{c^{\prime}})\bm{E}_{c}^{(-)}(\bm{r})\cdot\left(\overleftrightarrow{\varepsilon}[s_{c^{\prime}},\bm{\mathcal{M}}_{S}]\right)\cdot\bm{E}_{c^{\prime}}^{(+)}(\bm{r})e^{-i(s_{c^{\prime}}-s_{c})t}\\
 & \quad+(-is_{c^{\prime}})\bm{E}_{c}^{(+)}(\bm{r})\cdot\left(\overleftrightarrow{\varepsilon}[s_{c^{\prime}},\bm{\mathcal{M}}_{S}]\right)\cdot\bm{E}_{c^{\prime}}^{(+)}(\bm{r})e^{-i(s_{c^{\prime}}+s_{c})t}+{\rm c.c.}\Big],
\end{aligned}
\end{equation}
and
\begin{equation}
\begin{aligned}
w_{e,\delta}(t) & =\frac{\varepsilon_{0}}{4}\int_{-\infty}^{t}dt^{\prime}\sum_{c,c^{\prime}}\Big[-i(s_{c^{\prime}}+s_{m})\bm{E}_{c}^{(-)}(\bm{r})\cdot\left(\overleftrightarrow{\varepsilon}_{+}[s_{c^{\prime}},\delta\bm{\mathcal{M}}_{+}]\right)\cdot\bm{E}_{c^{\prime}}^{(+)}(\bm{r})e^{-i(s_{c^{\prime}}-s_{c}+s_{m})t}\\
 & \quad-i(s_{c^{\prime}}-s_{m})\bm{E}_{c}^{(-)}(\bm{r})\cdot\left(\overleftrightarrow{\varepsilon}_{-}[s_{c^{\prime}},\delta\bm{\mathcal{M}}_{-}]\right)\cdot\bm{E}_{c^{\prime}}^{(+)}(\bm{r})e^{-i(s_{c^{\prime}}-s_{c}-s_{m})t}\\
 & \quad-i(s_{c^{\prime}}+s_{m})\bm{E}_{c}^{(+)}(\bm{r})\cdot\left(\overleftrightarrow{\varepsilon}_{+}[s_{c^{\prime}},\delta\bm{\mathcal{M}}_{+}]\right)\cdot\bm{E}_{c^{\prime}}^{(+)}(\bm{r})e^{-i(s_{c^{\prime}}+s_{c}+s_{m})t}\\
 & \quad-i(s_{c^{\prime}}-s_{m})\bm{E}_{c}^{(+)}(\bm{r})\cdot\left(\overleftrightarrow{\varepsilon}_{-}[s_{c^{\prime}},\delta\bm{\mathcal{M}}_{-}]\right)\cdot\bm{E}_{c^{\prime}}^{(+)}(\bm{r})e^{-i(s_{c^{\prime}}+s_{c}-s_{m})t}+{\rm c.c}\Big].
\end{aligned}
\end{equation}
For $w_{e,S}(t)$ we use the same arguments as the monochromatic case, which assumes a RWA for the term $\propto e^{-i(s_{c^{\prime}}+s_{c})t}$, and yields
\begin{equation}
\begin{aligned}
u_{e,S} =\frac{\varepsilon_{0}}{4}\sum_{c}\bm{E}_{c}^{(-)}(\bm{r},t)\cdot\partial_{\omega_{c}}\left(\omega_{c}\overleftrightarrow{\varepsilon}[\omega_{c},\bm{\mathcal{M}}_{S}]\right)\cdot\bm{E}_{c}^{(+)}(\bm{r},t).
\end{aligned}
\end{equation}

For the term $w_{e,m}(t)$ we notice that the average over a period of oscillation imposes a triple-resonance condition: the only terms that are not discarded in the RWA are the ones $\propto e^{-i(s_{c}-s_{c^{\prime}}+s_{m})t}$ if $\omega_{c}-\omega_{c^{\prime}}\thickapprox\omega_{m}$, and the terms $\propto e^{-i(s_{c^{\prime}}-s_{c}+s_{m})t}$ if $\omega_{c^{\prime}}-\omega_{c}\thickapprox\omega_{m}$, that is, the frequency difference between mode $c$ and mode $c^{\prime}$ needs to match a magnon frequency. When the modes are degenerate, since the decomposition of the permittivity is linear in the magnetization and since we have used the RWA, we can recombine the terms of the decomposition, which will yield Eq.~\eqref{eq:EnergyDensityDispersion}. In that case, the energy mismatch is compensated by an external drive detuned by the magnon frequency. Assuming that the damping coefficients of all the modes are equal, we get the energy density
\begin{equation}
\begin{aligned}
u_{e,\delta} & =\frac{\varepsilon_{0}}{4}\int_{-\infty}^{t}dt^{\prime}\sum_{\omega_{c}>\omega_{c^{\prime}}}\Big[-is_{c}\bm{E}_{c}^{(-)}(\bm{r})\cdot\left(\overleftrightarrow{\varepsilon}_{+}[s_{c^{\prime}},\delta\bm{\mathcal{M}}_{+}]\right)\cdot\bm{E}_{c^{\prime}}^{(+)}(\bm{r})\\
 & \quad-is_{c^{\prime}}\bm{E}_{c^{\prime}}^{(-)}(\bm{r})\cdot\left(\overleftrightarrow{\varepsilon}_{-}[s_{c},\delta\bm{\mathcal{M}}_{-}]\right)\cdot\bm{E}_{c}^{(+)}(\bm{r})+{\rm c.c.}\Big]e^{2\alpha t}.
\end{aligned}
\end{equation}
The sum in the previous expression indicates that the indexes are organized such that $\omega_{c}>\omega_{c^{\prime}}$. Finally, we follow the same procedure for the monochromatic field, expanding $\overleftrightarrow{\varepsilon}_{\pm}[s_{c},\delta\bm{\mathcal{M}}_{\pm}]$ up to first order in $\alpha$ and using that $\omega_m \ll \omega_{c, c^\prime}$, we get after the time integration
\begin{equation}
\begin{aligned}
u_{e,\delta} & =\frac{\varepsilon_{0}}{4}\sum_{\omega_{c}>\omega_{c^{\prime}}}\Big[\bm{E}_{c}^{(-)}(\bm{r},t)\cdot\partial_{\omega_{c^{\prime}}}\left(\omega_{c^{\prime}}\overleftrightarrow{\varepsilon}_{+}[\omega_{c^{\prime}},\delta\bm{\mathcal{M}}_{+}]\right)\cdot\bm{E}_{c^{\prime}}^{(+)}(\bm{r},t)\\
 & \quad+\bm{E}_{c}^{(-)}(\bm{r},t)\cdot\partial_{\omega_{c}}\left(\omega_{c}\overleftrightarrow{\varepsilon}_{+}[\omega_{c},\delta\bm{\mathcal{M}}_{+}]\right)\cdot\bm{E}_{c^{\prime}}^{(+)}(\bm{r},t)+{\rm c.c.\Big]}.
\end{aligned}
\end{equation}
The total energy density is therefore given by
\begin{equation}
\begin{aligned}
u & =\frac{\varepsilon_{0}}{4}\sum_{c}\bm{E}_{c}^{(-)}(\bm{r},t)\cdot\partial_{\omega_{c}}\left(\omega_{c}\overleftrightarrow{\varepsilon}_{S}[\omega_{c},\bm{\mathcal{M}}_{S}]\right)\cdot\bm{E}_{c}^{(+)}(\bm{r},t)\\
 & +\frac{1}{4\mu_{0}}\sum_{c}\bm{B}_{c}^{(-)}(\bm{r},t)\cdot\bm{B}_{c}^{(+)}(\bm{r},t)+\sum_{\omega_{c}<\omega_{c^{\prime}}}u_{cc^{\prime}}\label{eq:FullEnergyDensity}
\end{aligned}
\end{equation}
where
\begin{equation}
\begin{aligned}
u_{cc^{\prime}} & =\frac{\varepsilon_{0}}{4}\Big[\bm{E}_{c}^{(-)}(\bm{r},t)\cdot\partial_{\omega_{c^{\prime}}}\left(\omega_{c^{\prime}}\overleftrightarrow{\varepsilon}_{+}[\omega_{c^{\prime}},\delta\bm{\mathcal{M}}_{+}]\right)\cdot\bm{E}_{c^{\prime}}^{(+)}(\bm{r},t)\\
 & \quad\quad\quad\quad+\bm{E}_{c}^{(-)}(\bm{r},t)\cdot\partial_{\omega_{c}}\left(\omega_{c}\overleftrightarrow{\varepsilon}_{+}[\omega_{c},\delta\bm{\mathcal{M}}_{+}]\right)\cdot\bm{E}_{c^{\prime}}^{(+)}(\bm{r},t)+{\rm c.c.\Big]}.
\end{aligned}
\end{equation}
\end{widetext}

Note that in deriving the expressions for the energy density Eqs.~\eqref{eq:EnergyDensityDispersion} and \eqref{eq:FullEnergyDensity} we employed the RWA, which can fail in the strong coupling regime  \cite{Diaz_2019_Ultrastrong}. This is usually the case when the matter excitations and the electromagnetic field have almost matching frequencies and the coupling rate becomes comparable to such frequencies. In the present case, the magnon typical frequency is far off-resonant from the photon frequency, such that even if the coupling between photons and magnons is comparable with the magnon frequency, the RWA is valid.

\subsection{Quantization of the electromagnetic field\label{subsec:Quantization-of-the}}

In order to quantize the energy density obtained in last section, we first decompose the electric field as
\begin{equation}
\begin{aligned}
\bm{E}(\bm{r},t) & =\sum_{c,l}\bm{E}_{c,l}(\bm{r},t)=\sum_{c,l}{\rm Re}\left[C_{c,l}\alpha_{c,l}(t)\bm{F}_{c,l}(\bm{r})\right] \\
&=\sum_{c,l}{\rm Re}\left[C_{c,l}\alpha_{c,l}(0)\bm{F}_{c,l}(\bm{r})e^{-i\omega_{c}t}\right],\label{eq:ModeDecomposition}
\end{aligned}
\end{equation}
with the indexes $l$ distinguishes degenerate modes (e.g. degenerate polarization states), and $\bm{F}_{c,l}(\bm{r})$ are the corresponding mode functions satisfying
\begin{equation}
\begin{aligned}
&\bm{\nabla}\cdot\bm{F}_{c,l}(\bm{r})  =0,\\
&\bm{\nabla}\times\bm{\nabla}\times\bm{F}_{c,l}(\bm{r})-\frac{\omega_{c}^{2}}{c^{2}}\overleftrightarrow{\varepsilon}[\omega_{c},\bm{\mathcal{M}}]\bm{F}_{c,l}(\bm{r})  =0,
\end{aligned}
\end{equation}
together with the corresponding boundary conditions. The mode functions $\bm{F}_{c,l}$ are also a function of the saturation magnetization. In magnetic dielectric optical cavities, the magnetization can break degeneracies, e.g., of whispering gallery modes in spherical cavities \cite{Ford_1978_scattering_absorption}, which can be exploited to couple magnons and photons, for example, in spherical cavity geometries \cite{Almpanis_2020_Spherical_optomagnonic}. For simplifying the notation, we omit the dependence of the mode functions on the magnetization. The induction field $\bm{B}(\bm{r},t)$ is given from Maxwell's equations by 
\begin{equation}
\bm{B}(\bm{r},t)=\sum_{c,l}{\rm Re}\left[-\frac{i}{\mu_{0}\omega_{c}}C_{c,l}\alpha_{c,l}(t)\bm{\nabla}\times\bm{F}_{c,l}(\bm{r})\right].
\end{equation}

For a specific mode frequency $\omega_{c}$, the energy density given by Eq.~\eqref{eq:EnergyDensityDispersion} can be decomposed as
\begin{equation}
u_{\omega_{c}}  =u_{\omega_{c}}^{({\rm D})}+u_{\omega_{c}}^{({\rm ND})},
\end{equation}
with
\begin{equation}
\begin{aligned}
u_{\omega_{c}}^{({\rm D})} & =\frac{1}{4}\sum_{l}\Big[\varepsilon_{0}\bm{E}_{c,l}^{(-)}\cdot\partial_{\omega_{c}}\left(\omega_{c}\overleftrightarrow{\varepsilon}[\omega_{c},\bm{\mathcal{M}}_{S}]\right)\cdot\bm{E}_{c,l}^{(+)} \\
&+\frac{1}{\mu_{0}}\bm{B}_{c,l}^{(-)}\cdot\bm{B}_{c,l}^{(+)}\Big],\\
u_{\omega_{c}}^{({\rm ND})} & =\frac{1}{4}\sum_{l\neq l^{\prime}}\Big[\varepsilon_{0}\bm{E}_{c,l}^{(-)}\cdot\partial_{\omega_{c}}\left(\omega_{c}\overleftrightarrow{\varepsilon}[\omega_{c},\bm{\mathcal{M}}]\right)\cdot\bm{E}_{c,l^{\prime}}^{(+)}\\
&+\frac{1}{\mu_{0}}\bm{B}_{c,l}^{(-)}\cdot\bm{B}_{c,l^{\prime}}^{(+)}\Big].
\label{eq:DecompositionEnergy}
\end{aligned}
\end{equation}
The first term $u_{\omega_{c}}^{({\rm D})}$, called here the diagonal term, does not couple different modes, and upon quantization yields the usual Hamiltonian of the electromagnetic field modes in terms of harmonic oscillators. To include dispersion in the quantization, we adopt the procedure by Milonni \cite{Milonni_1995_Field_Quantization,Milonni_2003_quantized_field,Raymer_2020_Quantum_theory}, which consists in taking the factors $C_{c,l}$ in Eq.~\eqref{eq:ModeDecomposition} and Eq.~\eqref{eq:EnergyDensityDispersion}, such that $u_{\omega_{c}}^{({\rm D})}$ yields a Hamiltonian of independent harmonic oscillators for each pair of mode indexes $c,l$. Those factors will depend on derivatives of the frequency-dependent permittivity and on the saturation magnetization. The second term in Eq.~\eqref{eq:DecompositionEnergy} $u_{\omega_{c}}^{({\rm ND})}$, which we call non-diagonal term, yields the magnon-photon coupling Hamiltonian with a coupling constant that is frequency-dependent and depends on derivatives of the elements of the permittivity tensor describing the effects of dispersion.

From Eq.~\eqref{eq:ModeDecomposition}, we obtain
\begin{equation}
U_{\omega_{c}}^{({\rm D})}  =\frac{1}{4}\sum_{l}\vert C_{c,l}\vert^{2}\vert\alpha_{c,l}\vert^{2}\mathcal{I}_{c,ll}[\omega_{c},\bm{\mathcal{M}}_{S}],
\label{eq:BigUD}
\end{equation}
where
\begin{equation}
\begin{aligned}
\mathcal{I}_{c,ll}[\omega_{c},\bm{\mathcal{M}}_{S}] & =\int d^{3}\bm{r}\Big\{\varepsilon_{0}\bm{F}_{l}^{*}\cdot\partial_{\omega_{c}}\left(\omega_{c}\overleftrightarrow{\varepsilon}[\omega_{c},\bm{\mathcal{M}}_{S}]\right)\cdot\bm{F}_{l} \\
 & \,\,\,\,\,\,\,\,\,\,+\frac{1}{\mu_{0}\omega_{c}^{2}}\vert \nabla\times\bm{F}_{l}\vert^2 \Big\}.
\end{aligned}
\end{equation}
By taking
\begin{equation}
C_{c,l}=\sqrt{2/\mathcal{I}_{c,ll}[\omega_{c},\bm{\mathcal{M}}_{S}]},\label{eq:NormConstant}
\end{equation}
equation \eqref{eq:BigUD} will be given by $U_{\omega_{c}}^{({\rm D})}=\sum_{l}\vert\alpha_{c,l}\vert^{2}/2$. We can then define the associated canonical momentum $p_{c,l}$ and position $q_{c,l}$ through $\alpha_{c,l}=p_{c,l}-i\omega_{c}q_{c,l}$, such that, the canonical equations of motion $\dot{p}_{c,l}=-\omega_{c}^{2}q_{c,l}$ and $\dot{q}=p$ are satisfied. Quantization is performed by promoting the canonical coordinates to operators satisfying canonical commutation relations $[\hat{q}_{c,l},\hat{p}_{c^{\prime},l^{\prime}}]=i\hbar\delta_{c,c^{\prime}}\delta_{l,l^{\prime}}$. The usual formulation in terms of creation and annihilation operators is obtained by writing $\hat{p}_{c,l}=(\hbar\omega_{c}/2)^{1/2}(\hat{a}_{c,l}+\hat{a}_{c,l}^{\dagger})$ and $\hat{q}_{c,l}=i(\hbar/2\omega_{c})^{1/2}(\hat{a}_{c,l}-\hat{a}_{c,l}^{\dagger})$ with $[\hat{a}_{c,l},\hat{a}_{c^{\prime},l^{\prime}}^{\dagger}]=\delta_{c,c^{\prime}}\delta_{l,l^{\prime}}$. Therefore, quantization yields $\alpha_{c,l}\rightarrow\sqrt{2\hbar\omega_{c}}\hat{a}_{c,l}$ and the Hamiltonian $U_{\omega_{c}}^{({\rm D})}\rightarrow\hat{H}_{\omega_{c}}=\sum_{l}\hbar\omega_{c}\hat{a}_{c,l}^{\dagger}\hat{a}_{c,l}$ besides the zero point energy term. The quantized electric field is therefore
\begin{equation}
\begin{aligned}
\hat{\bm{E}}(\bm{r},t)=
\sum_{c,l}\sqrt{\frac{4\hbar\omega_{c}}{\mathcal{I}_{c,ll}[\omega_{c},\bm{\mathcal{M}}_{S}]}}\left[\hat{a}_{c,l}(t)\bm{F}_{c,l}(\bm{r})+\hat{a}_{c,l}^{\dagger}\bm{F}_{c,l}^{*}(\bm{r})\right],
\end{aligned}
\end{equation}
For frequency ranges in which absorption is negligible, the procedure reproduces quantization procedures formulated in terms of a suitable Lagrangian including interactions with matter fields that describe dispersion and losses \cite{Barnett_1996_Field_Commutation,Huttner_1992_Dispersion_and_Loss,Huttner_1992_Quantization, Drezet_2017_Quantizing}. 

We can now proceed and obtain the coupling Hamiltonian. For degenerate modes, we first substitute the mode expansion for the electromagnetic field in $u_{\omega_{c}}^{({\rm ND})}$ (see Eq.~\eqref{eq:DecompositionEnergy}) such that
\begin{equation}
U_{\omega_{c}}^{({\rm ND})}=\frac{1}{4}\sum_{l\neq l^{\prime}}\alpha_{c,l}^{*}\alpha_{c,l^{\prime}}C_{c,l}C_{c,l^{\prime}}\mathcal{I}_{c,ll^{\prime}}[\omega_{c},\bm{\mathcal{M}}],
\label{eq:EnerDenND}
\end{equation}
where 
\begin{equation}
\begin{aligned}
\mathcal{I}_{c,ll^{\prime}}[\omega_{c},\bm{\mathcal{M}}] & =\int d^{3}\bm{r}\Big\{\varepsilon_{0}\bm{F}_{c,l}^{*}\cdot\partial_{\omega_{c}}\left(\omega_{c}\overleftrightarrow{\varepsilon}[\omega_{c},\bm{\mathcal{M}}]\right)\cdot\bm{F}_{c,l^{\prime}} \\
 & \,\,\,\,\,\,\,\,\,\,+\frac{1}{\mu_{0}\omega_{c}^{2}}\left(\nabla\times\bm{F}_{c,l}^{*}\right)\cdot\left(\nabla\times\bm{F}_{c,l^{\prime}}\right)\Big\}.
\end{aligned}
\label{eq:OverDeg}
\end{equation}
As outlined above, the electromagnetic field modes are quantized by $\alpha_{c,l}\rightarrow\sqrt{2\hbar\omega_{c}}\hat{a}_{c,l}$, while the magnetization is promoted to an operator $\bm{\mathcal{M}}\rightarrow\hat{\bm{\mathcal{M}}}$. With this procedure, we obtain the quantized optomagnonic Hamiltonian
\begin{equation}
\hat{H}_{cl,cl^{\prime}}=\sum_{l\neq l^{\prime}}\frac{\hbar \omega_{c} \mathcal{I}_{c,ll^{\prime}}[\omega_{c},\hat{\bm{\mathcal{M}}}]}{\sqrt{\mathcal{I}_{c,l^{\prime}l^{\prime}}[\omega_{c},\mathcal{M}_{S}]\mathcal{I}_{c,ll}[\omega_{c},\mathcal{M}_{S}]}}\hat{a}_{l}^{\dagger}\hat{a}_{l^{\prime}}.\label{eq:IntDege}
\end{equation}
The specific form of the coupling depends on the overlap integral $\mathcal{I}_{c,ll^{\prime}}[\omega_{c},\hat{\bm{\mathcal{M}}}]$ which in turn is given in terms of the magnetization-dependent mode functions $\bm{F}_{c,l}$ and on the derivatives of the permittivity tensor.

The coupling between different non-degenerate modes is obtained from $u_{cc^{\prime}}$. Specifically, the mode decomposition yields the energy
\begin{equation}
U_{cc^{\prime}}=\frac{1}{4}\sum_{l,l^{\prime}}\left[\alpha_{c,l}^{*}\alpha_{c^{\prime},l^{\prime}}C_{c,l}C_{c^{\prime},l^{\prime}}\mathcal{I}_{cl,c^{\prime}l^{\prime}}[\omega_{c},\omega_{c^{\prime}},\delta\bm{\mathcal{M}}_{+}]+{\rm c.c.}\right],
\label{eq:EnerDenNonDeg}
\end{equation}
where the overlap integral takes the form
\begin{widetext}
\begin{equation}
\mathcal{I}_{cl,c^{\prime}l^{\prime}}[\omega_{c},\omega_{c^{\prime}},\delta\bm{\mathcal{M}}_{+}]  =\varepsilon_{0}\int d^{3}\bm{r}\Big\{\bm{F}_{c,l}^{*}\cdot\partial_{\omega_{c^{\prime}}}\left(\omega_{c^{\prime}}\overleftrightarrow{\varepsilon}_{+}[\omega_{c^{\prime}},\delta\bm{\mathcal{M}}_{+}]\right)\cdot\bm{F}_{c^{\prime},l^{\prime}} +\bm{F}_{c,l}^{*}\cdot\partial_{\omega_{c}}\left(\omega_{c}\overleftrightarrow{\varepsilon}_{+}[\omega_{c},\delta\bm{\mathcal{M}}_{+}]\right)\cdot\bm{F}_{c^{\prime},l^{\prime}}\Big\}.\label{eq:OverNonDeg}
\end{equation}
The magnetization fluctuations are then quantized by $\delta\bm{\mathcal{M}}_{+}\rightarrow\delta\hat{\bm{\mathcal{M}}_{+}}$, such that $U_{cc^{\prime}}$ yields the interacting Hamiltonian
\begin{equation}
\hat{H}_{cl,c^{\prime}l^{\prime}}=\hbar\sum_{\omega_{c}>\omega_{c^{\prime}}}\left[\sqrt{\frac{\omega_{c}\omega_{c^{\prime}}}{\mathcal{I}_{c^{\prime},l^{\prime}l^{\prime}}[\omega_{c},\mathcal{M}_{S}]\mathcal{I}_{c,ll}[\omega_{c},\mathcal{M}_{S}]}}\mathcal{I}_{cl,c^{\prime}l^{\prime}}[\omega_{c},\omega_{c^{\prime}},\bm{e}_{\mathcal{\delta\mathcal{M}_{+}}}]\delta\hat{\mathcal{M}}_{+}\hat{a}_{c,l}^{\dagger}\hat{a}_{c^{\prime},l^{\prime}}+{\rm h.c.}\right],\label{eq:IntNonDeg}
\end{equation}
\end{widetext}
where we have used that $\mathcal{I}_{cl,c^{\prime}l^{\prime}}[\omega_{c},\omega_{c^{\prime}},\delta\bm{\mathcal{M}}_{+}]$ is linear in $\delta\bm{\mathcal{M}}_{+}$, and defined $\bm{e}_{\mathcal{\delta\mathcal{M}_{+}}}=\delta\bm{\mathcal{M}}_{+}/\vert\delta\bm{\mathcal{M}}_{+}\vert$. Within the Holstein-Primakoff transformation up to first order, $\delta\hat{\mathcal{M}}{}_{+}$ describes the annihilation of a magnon, and therefore $\hat{H}_{cl,c^{\prime}l^{\prime}}$ describes the process in which a photon in the mode $(c^{\prime},l^{\prime})$ with frequency $\omega_{c^{\prime}}$ annihilates a magnon with frequency $\omega_{m}$ creating a photon in mode $(c,l)$ with frequency $\omega_{c}=\omega_{c^{\prime}}+\omega_{m}$. In the degenerate case, the magnon is created provided that an external drive is detuned from the optical mode's resonance.

\subsection{Plane-wave like modes: Faraday and Voigt configurations\label{subsec:Plane-wave-like-modes:}}

To obtain the coupling and its explicit dependence on the functions $\varepsilon[\omega]$ and $\mathcal{F}[\omega]$, we need to specifythe mode functions $\bm{F}_{c,l}$ and compute the integrals $\mathcal{I}_{c,ll^{\prime}}[\omega_{c},\bm{\mathcal{M}}]$ and $\mathcal{I}_{cl,c^{\prime}l^{\prime}}[\omega_{c},\omega_{c^{\prime}},\bm{e}_{\mathcal{\delta\mathcal{M}_{+}}}]$. We first consider plane-wave like modes, such that $\bm{F}_{c,l}(\bm{r})\propto e^{i\bm{k}_{c,l}[\omega_{c}]\cdot\bm{r}}$. The modes of a Fabry-P\'{e}rot optomagnonical cavity can then described by considering perfect conducting boundary conditions confining the electromagnetic field along one dimension. For $\bm{F}_{c,l}(\bm{r})=e^{i\bm{k}_{c,l}[\omega_{c}]\cdot\bm{r}}\bm{f}_{c,l}$ from Maxwell's equation we obtain
\begin{equation}
\left[\overleftrightarrow{\mathcal{K}}_{c,l}\cdot\overleftrightarrow{\mathcal{K}}_{c,l}+\frac{\omega_{c}^{2}}{c^{2}}\overleftrightarrow{\varepsilon}[\omega_{c},\bm{\mathcal{M}}]\right]\cdot\bm{\bm{f}}_{l}=0,\label{eq:ElectricPlane}
\end{equation}
where
\[
\overleftrightarrow{\mathcal{K}}_{c,l}=\left[\begin{array}{ccc}
0 & -k_{c,l;z} & k_{c,l;y}\\
k_{c,l;z} & 0 & -k_{c,l;x}\\
-k_{c,l;y} & k_{c,l;x} & 0
\end{array}\right].
\]
Thus the wave vector $\bm{k}_{c,l}$ satisfy the Fresnel's equation
\begin{equation}
{\rm det}\left[\overleftrightarrow{\mathcal{K}}_{c,l}\cdot\overleftrightarrow{\mathcal{K}}_{c,l}+\frac{\omega_{c}^{2}}{c^{2}}\overleftrightarrow{\varepsilon}[\omega_{c},\bm{\mathcal{M}}]\right]=0.\label{eq:Fresnel}
\end{equation}
The solutions of the above equation give the relation between the wave vector and frequency $k_{c,l}[\omega_{c}]$ and, once the boundary conditions are applied, the discrete values of $k_{c,l}$, and the corresponding allowed frequencies are obtained.

\begin{figure}
\centering{}\includegraphics[width=1.\columnwidth]{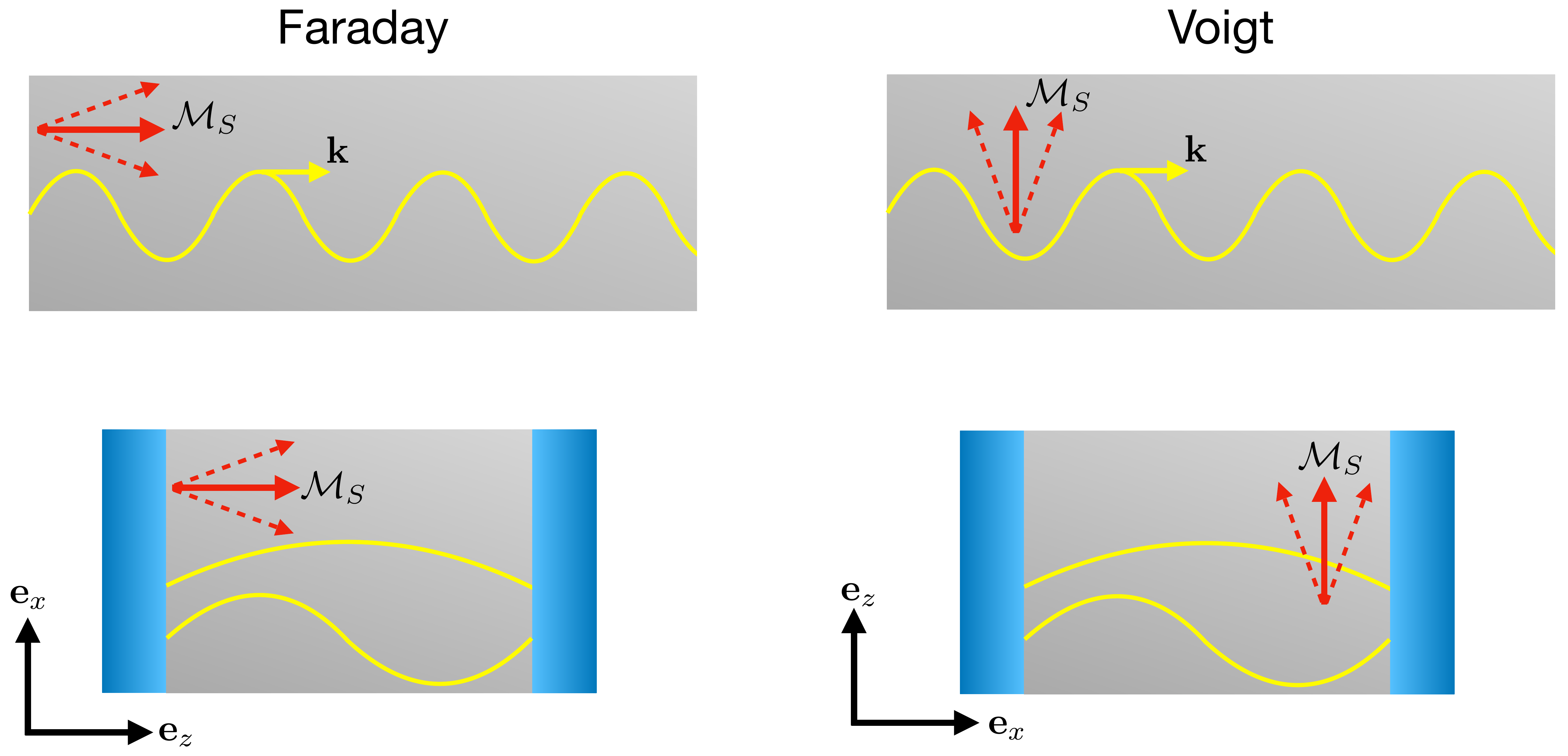}\caption{Faraday and Voigt configurations. In the Faraday configuration, the
wave vector $\bm{k}_{c,l}$ is parallel to the saturation magnetization,
while in the Voigt configuration, the wave vector is perpendicular
to the saturation magnetization. We also consider the corresponding
Fabry-P\'{e}rot cavities by imposing perfect conducting boundary conditions.}
\label{Configurations}
\end{figure}

We focus on two configurations: the Faraday configuration, for which $\bm{k}_{c,l}[\omega_{c}]$ is parallel to the saturation magnetization, and the Voigt configuration, for which $\bm{k}_{c,l}[\omega_{c}]$ is perpendicular to the saturation magnetization, as depicted in Fig. ~\ref{Configurations}. Since we are considering $\bm{\mathcal{M}}_{S}=\mathcal{M}_{S}\bm{e}_{z}$, for the Faraday configuration $\bm{k}_{c,l}[\omega_{c}]\parallel\bm{e}_{z}$, while for the Voigt configuration we take $\bm{k}_{c,l}[\omega_{c}]\parallel\bm{e}_{x}$.

\subsubsection{Faraday configuration}

In the Faraday configuration the Fresnel's equation yields two wave-vectors, labeled with indexes $l=\pm$ and given by 
\begin{equation}
k_{c,\pm}^{2}=\frac{\omega_{c}^{2}}{c^{2}}\left[\varepsilon[\omega_{c}]\pm\left|\mathcal{F}[\omega_{c}]\right|\mathcal{M}_{S}\right],\label{eq:FaradayK}
\end{equation}
The magnetization-dependent mode functions obtained from the Fresnel's equation corresponding to $k_{c,\pm}$ are
\begin{equation}
\begin{aligned}
\bm{F}_{c,\pm}(\bm{r}) & =\bm{f}_{c,\pm}\frac{\left(\mathbb{A}_{c,\pm}e^{ik_{c,\pm}[\omega_{c}]z}+\mathbb{B}_{c,\pm}e^{-ik_{c,\pm}[\omega_{c}]z}\right)}{\sqrt{V}}, \\
\bm{f}_{c,\pm} & =\bm{e}_{x}\mp i{\rm sign}\left[\mathcal{F}[\omega_{c}]\right]\bm{e}_{y} \\
&\mp\frac{\mathcal{F}[\omega_{c}]}{\varepsilon[\omega_{c}]}\left(\pm i\mathcal{M}_{y}-{\rm sign}\left[\mathcal{F}[\omega_{c}]\right]\mathcal{M}_{x}\right)\bm{e}_{z},
\label{eq:FaradayModes}
\end{aligned}
\end{equation}
where we have considered only terms up to linear order in $\mathcal{M}_{x,y}$ and took $\mathcal{M}_{z}\sim\mathcal{M}_{S}$. For $\mathcal{M}_{x,y}\rightarrow0$, those modes are circularly polarized with mode-functions in the $xy$ plane. For plane waves propagating in the $\bm{e}_{z}$ direction in an unbounded medium, we take $\mathbb{A}_{\pm}=1$ and $\mathbb{B}_{\pm}=0$, for which the normalization constant Eq.~\eqref{eq:NormConstant} is given by
\begin{equation}
\begin{aligned}
C_{c,\pm}[\omega_{c}]=\Big[ &\varepsilon_{0}\Big( \vert\mathbb{K}_{\pm}[\omega_{c}]\vert^{2}+\partial_{\omega_{c}}\left( \omega_{c} \varepsilon[\omega_{c}] \right)\\
&\pm{\rm sign}[\mathcal{F}[\omega_{c}]]\partial_{\omega_{c}}\left(\omega_{c}\mathcal{F}[\omega_{c}]\right)\mathcal{M}_{S}\Big)/2 \Big]^{-1/2},
\label{eq:NormFaraday-1}
\end{aligned}
\end{equation}
where
\begin{align}
\mathbb{K}_{\pm}[\omega_{c}] & =\left[\varepsilon[\omega_{c}]\pm\mathcal{M}_{S}\left|\mathcal{F}[\omega_{c}]\right|\right]^{1/2}.
\end{align}

\subsubsection{Voigt configuration}

For the Voigt configuration $\bm{k}_{c,l}=k_{c,l}\bm{e}_{x}$ and, similarly to the Faraday case, we obtain two wave-vectors labeled with $l=\pm$ given by
\begin{equation}
\begin{aligned}
k_{c,+}^{2}[\omega_{c}]&=\frac{\omega_{c}^{2}}{c^{2}}\varepsilon[\omega_{c}], \\
k_{c,-}^{2}[\omega_{c}]&=\frac{\omega_{c}^{2}}{c^{2}}\left(\varepsilon[\omega_{c}]-\frac{\mathcal{F}^{2}[\omega_{c}]\mathcal{M}_{S}^{2}}{\varepsilon[\omega_{c}]}\right),
\label{eq:VoigtK}
\end{aligned}
\end{equation}
with mode functions
\begin{widetext}
\begin{equation}
\begin{aligned}
\bm{F}_{c,\pm}(\bm{r}) & =\bm{f}_{c,\pm}\frac{\mathbb{A}_{c,\pm}e^{ik_{c,\pm}[\omega_{c}]x}+\mathbb{B}_{c,\pm}e^{-ik_{c,\pm}[\omega_{c}]x}}{\sqrt{V}},\\
\bm{f}_{c,+} & =\bm{e}_{z}+\frac{\mathcal{M}_{x}}{\mathcal{M}_{S}}\bm{e}_{x}+\left(\frac{i\varepsilon[\omega_{c}]\mathcal{M}_{x}}{\mathcal{F}[\omega_c]\mathcal{M}_{S}^{2}}+\frac{\mathcal{M}_{y}}{\mathcal{M}_{S}}\right)\bm{e}_{y}\, \\
\bm{f}_{c,-} & =\bm{e}_{y}-i\frac{\mathcal{F}[\omega_c]}{\varepsilon[\omega_c]}\mathcal{M}_{S}\bm{e}_{x}+\left(\frac{i\varepsilon[\omega_c]\mathcal{M}_{x}}{\mathcal{F}[\omega_c]\mathcal{M}_{S}^{2}}-\frac{\mathcal{M}_{y}}{\mathcal{M}_{S}}\right)\bm{e}_{z}.
\label{eq:VoigtModes}
\end{aligned}
\end{equation}
We notice that the mode $+$ has a wave-vector independent of the magnetization and it is imaginary in the regions were $\varepsilon[\omega_{c}]<0$. Otherwise, the mode $-$ has an imaginary wave-vector for frequencies such that $\varepsilon^{2}[\omega_{c}]<\mathcal{F}^{2}[\omega_{c}]\mathcal{M}_{S}^{2}$. We restrict our analysis to the case $\varepsilon[\omega_{c}]>0$. For the case of plane waves propagating in a bulk medium along $\bm{e}_{x}$, we take $\mathbb{A}_{c,\pm}=1$ and $\mathbb{B}_{c,\pm}=0$. In this case, the normalization factors are given by

\begin{equation}
\begin{aligned}
C_{c,+}[\omega_{c}] & =\left[\varepsilon_{0}\left(\partial_{\omega_{c}}\left(\omega_{c}\varepsilon[\omega_{c}]\right)+\varepsilon[\omega_{c}]\right)/2\right]^{-1/2},\\
C_{c,-}[\omega_{c}] & =\Big[\varepsilon_{0}\Big(\left(1+\frac{\mathcal{F}^{2}[\omega_{c}]\mathcal{M}_{S}^{2}}{\varepsilon^{2}[\omega_{c}]}\right)\partial_{\omega_{c}}(\omega_{c}\varepsilon[\omega_{c}])\\
&\quad -2\frac{\mathcal{F}[\omega_{c}]}{\varepsilon[\omega_{c}]}\partial_{\omega_{c}}(\omega_{c}\mathcal{F}[\omega_{c}])\mathcal{M}_{S}^{2}+\vert\varepsilon[\omega_{c}]-\frac{\mathcal{F}^{2}[\omega_{c}]\mathcal{M}_{S}^{2}}{\varepsilon[\omega_{c}]}\vert \Big)/2\Big]^{-1/2}.
\end{aligned}
\end{equation}
\end{widetext}
We notice that $C_{c,+}[\omega_{c}]$ has the same form as the normalization constant for plane-wave modes in a homogeneous dispersive materials with no permeability dispersion \cite{Lobet_2020_Fundamental}.

\subsubsection{Fabry-P\'{e}rot cavity\label{Cavity}}

The description of a Fabry-P\'{e}rot cavity requires the imposition of boundary conditions. We consider perfect conducting boundary conditions in the two different configurations: the Faraday (the magnetization is perpendicular to the cavity boundaries) and the Voigt configuration (the magnetization is parallel to the cavity boundary). Otherwise, all-dielectric cavities host quasi-normal modes \cite{Leung_1994_completeness,Mansuripur_2017_leaky_modes,Jakobsen_2018_leakymode}, that take into account leakage of the electromagnetic field from the dielectric to its surrounds. The proper quantization of quasi-normal modes is relevant for all-dielectric cavity quantum electrodynamics setups \cite{Franke_2019_Quantization_of_Quasinormal,Franke_2020_quantizedquasinormal}, which we do not discuss here.

For the cavity in the Faraday-like configuration, we take the wave-vector as $\bm{k}=\bm{q}+k_{c,l}\bm{e}_{z}$, where $\bm{q}=k_{x}\bm{e}_{x}+k_{y}\bm{e}_{y}$. From the Fresnel equations we obtain
\begin{equation}
\begin{aligned}
k_{c,\pm}^{2}[\omega_{c}]&=\frac{\omega_{c}^{2}}{c^{2}}\varepsilon[\omega_{c}]-q^{2} \\
&\pm\frac{1}{\varepsilon[\omega_{c}]}\sqrt{\frac{\omega_{c}^{2}}{c^{2}}\varepsilon[\omega_{c}]\mathcal{F}^{2}[\omega_{c}]\mathcal{M}_{S}^{2}\left(\frac{\omega_{c}^{2}}{c^{2}}\varepsilon[\omega_{c}]-q^{2}\right)},\label{eq:CavityFar}
\end{aligned}
\end{equation}
The normal modes of the cavity have thus the following form
\begin{equation}
\bm{F}_{c,\pm}(\bm{r})=\bm{f}_{c,\pm}[\bm{q}]\frac{e^{i\bm{q}\cdot\bm{\rho}}}{\sqrt{V}}\left(\mathbb{A}_{c,\pm}e^{-ik_{c,\pm}z}+\mathbb{B}_{c,\pm}e^{ik_{c,\pm}z}\right),
\end{equation}
where $\bm{q}=k_{x}\bm{e}_{x}+k_{y}\bm{e}_{y}$, $\bm{\rho}=x\bm{e}_{x}+y\bm{e}_{y}$ and $V$ is the volume of the cavity. The mode vectors $\bm{f}_{c,\pm}[\bm{q}]$ are given from Eq.~\eqref{eq:ElectricPlane} which for $q=0$ are given by Eq.~\eqref{eq:FaradayModes}. We consider perfect conducting boundary conditions $\bm{e}_{z}\times\bm{F}_{c,\pm}(\bm{r})=0$ and $\bm{e}_{z}\cdot\left(\varepsilon_{0}\overleftrightarrow{\varepsilon}[\omega_{c},\bm{\mathcal{M}}_{S}]\cdot\bm{F}_{c,\pm}(\bm{r})\right)=0$ at $z=0,L$. From the first condition at $z=0$ we get $\mathbb{A}_{c,\pm}=-\mathbb{B}_{c,\pm}$. Therefore
\begin{equation}
\bm{F}_{c,\pm}(\bm{r})=i\tilde{\mathbb{A}}_{c,\pm}\bm{f}_{c,\pm}[\bm{q}]\frac{e^{i\bm{q}\cdot\bm{\rho}}}{\sqrt{V}}\sin(k_{c,\pm}z),\label{eq:FaradayFP}
\end{equation}
where $\tilde{\mathbb{A}}_{c,\pm}$ is a redefined normalization constant that can be included in the normalization factor of the quantization procedure. The boundary condition at $z=L$ yields $k_{c,\pm}=2\pi n_{\pm}/L$ with $n_{\pm}$ integers, and Eq.~\eqref{eq:CavityFar} determines $\omega_{n_{\pm},\pm}[q]$ for each mode. The exact form of $\omega_{n_{\pm},\pm}[q]$ depends on the dispersion model.

Following the same procedure for the Voigt-like configuration cavity $\bm{k}=\bm{q}+k_{c,l}\bm{e}_{x}$, where now $\bm{q}=k_{z}\bm{e}_{z}+k_{y}\bm{e}_{y}$ gives
\begin{equation}
\begin{aligned}
k_{c,\pm}^{2}[\omega_{c}]&=\frac{\omega_{c}^{2}}{c^{2}}\varepsilon[\omega_{c}]-\frac{\omega_{c}^{2}\mathcal{F}^{2}[\omega_{c}]\mathcal{M}_{S}^{2}}{2c^{2}\varepsilon[\omega_{c}]}-q^{2} \\
&\pm\frac{1}{2}\sqrt{\frac{\omega_{c}^{2}\mathcal{F}^{2}[\omega_{c}]\mathcal{M}_{S}^{2}}{c^{2}\varepsilon^{2}[\omega_{c}]}\left(\frac{\omega_{c}^{2}\mathcal{F}^{2}[\omega_{c}]\mathcal{M}_{S}^{2}}{c^{2}}+4k_{z}^{2}\varepsilon[\omega_{c}]\right)},
\end{aligned}
\end{equation}
The boundary conditions are now $\bm{e}_{x}\times\bm{F}_{c,\pm}(\bm{r})=0$ and $\bm{e}_{x}\cdot\left(\varepsilon_{0}\overleftrightarrow{\varepsilon}[\omega_{c},\bm{\mathcal{M}}_{S}]\cdot\bm{F}_{c,\pm}(\bm{r})\right)=0$ at $x=0,L$, which gives the discrete allowed values for the mode frequencies $\omega_{n_{\pm},\pm}[q]$. The mode functions have a form similar to Eq.~\eqref{eq:FaradayFP}, being proportional to $\sin\left(k_{c,\pm}x\right)$ and with mode vectors $\bm{f}$ given, for $\bm{q}=\bm{0}$ by Eqs.~\eqref{eq:VoigtModes}.

Both for the Faraday and Voigt cavity configurations, the modes with $\bm{q}=0$ have normalization constants given by Eq.~\eqref{eq:NormConstant}, equal to those obtained for traveling plane waves. While the Faraday modes exhibit, up to first order in the saturation magnetization, perfect circular polarization (see Eq.~\eqref{eq:FaradayModes}), due to the circular magnetic birefringence one the Voigt mode labeled $(-)$ in Eqs. has a polarization component parallel to its wave vector. As we show in the following, due to the polarization of the Faraday modes, the Kittel-mode-mediated scattering between these modes is not allowed.

\subsection{Optomagnonic coupling Hamiltonian\label{subsec:Optomagnonic-coupling-Hamiltonia}}

With the mode functions and the quantized optical modes, we can obtain the magnon-photon interaction Hamiltonian from Eq.~\eqref{eq:IntDege} and Eq.~\eqref{eq:IntNonDeg}. We consider the case of plane waves or cavity modes with $\bm{q}=0$, since those have the same mode polarizations as the plane-waves (see Sec. \ref{Cavity}). Since in the previous section we have assumed that the saturation magnetization is polarized in the $\bm{e}_{z}$ direction, the fluctuations are entirely in the $xy$ plane. We therefore take $\delta\mathcal{M}_{\pm}=\mathcal{M}_{x}\mp i\mathcal{M}_{y}$.

For the Faraday configuration, we first notice that the second term in Eq.~\eqref{eq:OverDeg} yields no coupling since $\left(\nabla\times\bm{F}_{+}^{*}\right)\cdot\left(\nabla\times\bm{F}_{-}\right)=0$. Furthermore, up to first order in the magnetization's fluctuations
\begin{equation}
\bm{F}_{c^{\prime},-}^{*}(\bm{r})\cdot\partial_{\omega_{c}}(\omega_{c}\overleftrightarrow{\varepsilon}[\omega_{c},\bm{\mathcal{M}}])\cdot\bm{F}_{c,+}(\bm{r})=0,
\end{equation}
\\
and therefore $\mathcal{I}_{c,+-}[\omega_{c},\bm{\mathcal{M}}]=\mathcal{I}_{c,-+}[\omega_{c},\bm{\mathcal{M}}]=\mathcal{I}_{c+,c^{\prime}-}[\omega_{c},\omega_{c^{\prime}},\delta\bm{e}_{+}]=0$ and the magnon-mediated coupling between Faraday modes vanishes for both degenerate and non-degenerate modes.

In the Voigt configuration we also have $\left(\nabla\times\bm{F}_{-}^{*}\right)\cdot\left(\nabla\times\bm{F}_{+}\right)=0$. For degenerate modes we obtain
\begin{widetext}
\begin{equation}
\bm{F}_{-}^{*}\cdot\partial_{\omega_{c}}(\omega_{c}\overleftrightarrow{\varepsilon}[\omega_{c},\bm{\mathcal{M}}])\cdot\bm{F}_{+} =i\frac{e^{i(\bm{k}_{c,+}-\bm{k}_{c,-})\cdot\bm{r}}}{V}  \left[\frac{\mathcal{F}[\omega_{c}]}{\varepsilon[\omega_{c}]}\partial_{\omega_{c}}\left(\omega_{c}\varepsilon[\omega_{c}]\right)-\partial_{\omega_{c}}\left(\omega_{c}\mathcal{F}[\omega_{c}]\right)\right]\mathcal{M}_{x}.
\end{equation}
and thus
\begin{equation}
\mathcal{I}_{c,-+}[\omega_{c},\bm{\mathcal{M}}]=i\Xi\varepsilon_{0}\left[\frac{\mathcal{F}[\omega_{c}]}{\varepsilon[\omega_{c}]}\partial_{\omega_{c}}\left(\omega_{c}\varepsilon[\omega_{c}]\right)-\partial_{\omega_{c}}\left(\omega_{c}\mathcal{F}[\omega_{c}]\right)\right]\mathcal{M}_{x},
\end{equation}
\end{widetext}
where $\Xi$ is a mode overlap integral, which can be used as a generalization to the plane wave framework presented here. We follow the quantization procedure outlined above and consider the Holstein-Primakoff transformation up to linear order for the magnetization operator $\hat{\mathcal{M}}_{x}\approx\mathcal{M}_{{\rm ZPF}}(\hat{m}^{\dagger}+\hat{m})$, where $\mathcal{M}_{{\rm ZPF}}=\left(\gamma\hbar\mathcal{M}_{S}/2V\right)^{1/2}$ describes the zero point fluctuations of the magnetization given in terms of the volume of the magnet $V$ and of the gyromagnetic ratio $\gamma$, and $\hat{m}^{\dagger}$ ($\hat{m}$) is the magnon creation (annihilation) operator, satisfying bosonic commutation relations. The Hamiltonian obtained is given by
\begin{equation}
\hat{H}_{c,{\rm OM}}=i\hbar g_{{\rm Deg}}[\omega_{c}]\hat{a}_{c,+}^{\dagger}\hat{a}_{c,-}(\hat{m}^{\dagger}+\hat{m})+{\rm h.c.}\label{eq:OptoMagCoupDege}
\end{equation}
where the frequency-dependent coupling constant $g[\omega_{c}]$ reads
\begin{equation}
\begin{aligned}
g_{{\rm Deg}}[\omega_{c}] & =\frac{\omega_{c}\varepsilon_{0}}{2}\mathcal{M}_{{\rm ZPF}}\Xi C_{+}[\omega_{c}]C_{-}[\omega_{c}],\\
 & \times\left[\frac{\mathcal{F}[\omega_{c}]}{\varepsilon[\omega_{c}]}\partial_{\omega_{c}}\left(\omega_{c}\varepsilon[\omega_{c}]\right)-\partial_{\omega_{c}}\left(\omega_{c}\mathcal{F}[\omega_{c}]\right)\right],
 \label{eq:CoupVoiu}
\end{aligned}
\end{equation}
as also derived in \cite{Bittencourt_2021_ENZ_Lett}. For non-degenerate modes, we have two possible frequency configurations: (I) $\omega_{c,+}=\omega_{c^{\prime},-}+\omega_{m}$ and (II) $\omega_{c,-}=\omega_{c^{\prime},+}+\omega_{m}$. For (I), we get for the overlap integral Eq.~\eqref{eq:OverNonDeg}
\begin{widetext}
\begin{equation}
\mathcal{I}_{c+,c^{\prime}-}[\omega_{c,+},\omega_{c^{\prime},-},\delta\bm{e}_{+}]=i\Xi\varepsilon_{0}\frac{\partial_{\omega_{c,+}}\left(\omega_{c,+}\mathcal{F}[\omega_{c,+}]\right)+\partial_{\omega_{c^{\prime},-}}\left(\omega_{c^{\prime},-}\mathcal{F}[\omega_{c^{\prime},-}]\right)}{2}\left(\frac{\mathcal{F}[\omega_{c^{\prime},-}]\mathcal{M}_{S}}{\varepsilon[\omega_{c^{\prime},-}]}-1\right),
\end{equation}
Using the Holstein-Primakoff approximation $\hat{\mathcal{M}}_{+}=2\mathcal{M}_{{\rm ZPF}}\hat{m}$, and the Hamiltonian from Eq.~\eqref{eq:IntNonDeg} reads
\begin{equation}
\hat{H}_{\omega_{c,+}>\omega_{c^{\prime},-}}=i\hbar g[\omega_{c,+},\omega_{c^{\prime},-}]\hat{a}_{+}^{\dagger}\hat{a}_{-}\hat{m}+{\rm h.c.}\label{eq:NonDegHam01}
\end{equation}
with the coupling given by
\begin{equation}
\begin{aligned}
g[\omega_{c,+},\omega_{c^{\prime}-}] & =\frac{\varepsilon_{0}}{4}\Xi\mathcal{M}_{{\rm ZPF}}\sqrt{\omega_{c,+}\omega_{c^{\prime},-}}C_{c,+}[\omega_{c}]C_{c^{\prime},-}[\omega_{c^{\prime}}]\\
 & \times\left[\partial_{\omega_{c,+}}\left(\omega_{c,+}\mathcal{F}[\omega_{c,+}]\right)+\partial_{\omega_{c^{\prime},-}}\left(\omega_{c^{\prime},-}\mathcal{F}[\omega_{c^{\prime},-}]\right)\right]\left(\frac{\mathcal{F}[\omega_{c^{\prime},-}]\mathcal{M}_{S}}{\varepsilon[\omega_{c^{\prime},-}]}-1\right).
\label{eq:NonDegCoupl1}
\end{aligned}
\end{equation}
For (II), the overlap integral from Eq.~\eqref{eq:OverNonDeg} is
\begin{equation}
\mathcal{I}_{c-,c^{\prime}+}[\omega_{c,-},\omega_{c^{\prime},+},\delta\bm{e}_{+}]=i\Xi\varepsilon_{0}\frac{\partial_{\omega_{c,-}}\left(\omega_{c,-}\mathcal{F}[\omega_{c,-}]\right)+\partial_{\omega_{c^{\prime},+}}\left(\omega_{c^{\prime},+}\mathcal{F}[\omega_{c^{\prime},+}]\right)}{2}\left(\frac{\mathcal{F}[\omega_{c,-}]\mathcal{M}_{S}}{\varepsilon[\omega_{c,-}]}+1\right).
\end{equation}
The Hamiltonian Eq.~\eqref{eq:IntNonDeg} in this case reads 
\begin{equation}
\hat{H}_{\omega_{c,-}>\omega_{c^{\prime},+}}=i\hbar g[\omega_{c,-},\omega_{c^{\prime},+}]\hat{a}_{-}^{\dagger}\hat{a}_{+}\hat{m}+{\rm h.c.}\label{eq:NonDegHam02}
\end{equation}
where the coupling is
\begin{equation}
\begin{aligned}
g[\omega_{c,-},\omega_{c^{\prime},+}] & =\frac{\varepsilon_{0}}{4}\Xi\mathcal{M}_{{\rm ZPF}}\sqrt{\omega_{c,-}\omega_{c^{\prime},+}}C_{c,-}[\omega_{c}]C_{c^{\prime},+}[\omega_{c^{\prime}}]\\
 & \times\left[\partial_{\omega_{c,-}}\left(\omega_{c,-}\mathcal{F}[\omega_{c,-}]\right)+\partial_{\omega_{c^{\prime},+}}\left(\omega_{c^{\prime},+}\mathcal{F}[\omega_{c^{\prime},+}]\right)\right]\left(\frac{\mathcal{F}[\omega_{c,-}]\mathcal{M}_{S}}{\varepsilon[\omega_{c,-}]}+1\right).
 \label{eq:NonDegCoup2}
\end{aligned}
\end{equation}
\end{widetext}
Notice that both couplings have a similar form with a sign difference in the last term in brackets. The terms included in the permittivity tensor do not yield co-polarization coupling since $g_{\omega_{c,-}<\omega_{c^{\prime},-}}=g_{\omega_{c,+}<\omega_{c^{\prime},+}}=0$, although these can be generated by the terms associated with the Cotton-Mouton effect, which were not taken into account in our formalism.

\subsubsection{General properties of the optomagnonic coupling}

Without specifying the dispersion model, we can draw some general conclusions regarding the obtained optomagnonic coupling. We are particularly interested in the behavior of the coupling for frequencies near $\omega_{{\rm ENZ}}$.

For the degenerate case, we first notice that, as long as $\mathcal{F}^{2}[\omega_{c}]$ and $\partial_{\omega_{c}}(\mathcal{F}^{2}[\omega_{c}])$ are finite as $\omega_{c}\rightarrow\omega_{{\rm ENZ}}$ we obtain
\begin{equation}
{\rm lim}_{\omega_{c}\rightarrow\omega_{{\rm ENZ}}}\left[\frac{C_{-}[\omega_{c}]}{\varepsilon[\omega_{c}]}\right]=\sqrt{\frac{2}{\varepsilon_{0}\omega_{c}\mathcal{F}^{2}[\omega_{c}]\mathcal{M}_{S}^{2}\partial_{\omega_{c}}\left(\varepsilon[\omega_{c}]\right)}},
\end{equation}
which after some algebraic manipulations leads to \cite{Bittencourt_2021_ENZ_Lett}
\begin{equation}
{\rm lim}_{\omega_{c}\rightarrow\omega_{{\rm ENZ}}}\left[g_{{\rm Deg}}[\omega_{c}]\right]=\frac{\mathcal{M}_{{\rm ZPF}}}{\mathcal{M}_{S}}\Xi\omega_{{\rm ENZ}}\label{eq:EZCoupling}
\end{equation}
apart from an irrelevant ${\rm sign\left[\mathcal{F}[\omega_{{\rm ENZ}}]\right]}$. Therefore, irrespective of the dispersion model, at the ENZ frequency the coupling between magnons and photons is given by the simple relation of Eq.~\eqref{eq:EZCoupling} which depends on the ratio between the zero-point magnetic fluctuations and the saturation magnetization, the overlap integral and the ENZ frequency.

In the case of coupling between non-degenerate modes, we notice that the optomagnonic coupling for case (I) $\omega_{c,+}>\omega_{c^{\prime},-}$, given by Eq.~\eqref{eq:NonDegCoupl1}, vanishes if $\mathcal{F}[\omega_{c^{\prime},-}]\mathcal{M}_{S}=\varepsilon[\omega_{c^{\prime},-}]$, even though the energy conservation condition $\omega_{c,+}=\omega_{c^{\prime},-}+\omega_{m}$ is by assumption fulfilled. In this case, a photon in the mode $(c^{\prime},-)$ can not be \textit{upconverted} to a photon in $\left(c,+\right)$ by the absorption of a magnon. A complementary situation happens for case (II) $\omega_{c,-}>\omega_{c^{\prime},+}$ if $\mathcal{F}[\omega_{c,-}]\mathcal{M}_{S}=-\varepsilon[\omega_{c,-}]$: in this case the coupling in Eq.~\eqref{eq:NonDegCoup2} vanishes, and a photon in $(c,-)$ can not be \textit{downconverted} to a photon in $(c^{\prime},+)$ by the emission of a photon. For both cases, the frequencies at which the coupling vanishes corresponds to the frequency at which the wave-vector $k_{c,-}$ vanishes, given by (c.f. Eq.~\eqref{eq:VoigtK}):
\begin{equation}
\varepsilon[\omega_{c,-}]=\pm \mathcal{F}[\omega_{c,-}] \mathcal{M}_S.
\label{FreqSelectRules}
\end{equation}

We can understand the suppression of the coupling at these specific frequencies by inspecting the overlap integral Eq.~\eqref{eq:OverNonDeg}. We first notice that for the considered configuration with saturation magnetization along $\bm{e}_{z}$
\begin{equation}
\begin{aligned}
\partial_{\omega_{c^{\prime}}} &\Big( \omega_{c^{\prime}} \overleftrightarrow{\varepsilon}_{+}  [\omega_{c^{\prime}},\delta\bm{\mathcal{M}}_{+}] \Big)\\
&=\frac{\partial_{\omega_{c^{\prime}}}\left(\omega_{c^{\prime}}\mathcal{F}[\omega_{c^{\prime}}]\right)(\mathcal{M}_{x} - i \mathcal{M}_{y}  )}{\sqrt{2}}\left(\bm{e}_{z}\bm{e}_{+}-\bm{e}_{+}\bm{e}_{z}\right),\label{eq:DecompEpsPlus}
\end{aligned}
\end{equation}
where $\bm{e}_{\pm}=(\bm{e}_{x} \pm i \bm{e}_{y})/\sqrt{2}$ and a similar expression is valid for $\partial_{\omega_{c}}\left(\omega_{c}\overleftrightarrow{\varepsilon}_{+}[\omega_{c},\delta\bm{\mathcal{M}}_{+}]\right)$. The coupling between two non-degenerate plane wave-like modes $(c,l)$ and $(c^{\prime},l^{\prime})$, with $\omega_{c,l}>\omega_{c^{\prime},l^{\prime}}$ is thus proportional to $(\bm{f}_{c,l}^{*}\cdot\bm{e}_{z})(\bm{e}_{+}\cdot\bm{f}_{c^{\prime},l^{\prime}})-(\bm{f}_{c,l}^{*}\cdot\bm{e}_{+})(\bm{e}_{z}\cdot\bm{f}_{c^{\prime},l^{\prime}})$. This term does not vanish if either $\bm{f}_{c^{\prime},l^{\prime}}$ has a left-handed (LH) polarization component (a component $\parallel\bm{e}_{-}$) and $\bm{f}_{c,l}$ has a component $\parallel\bm{e}_{z}$, or when $\bm{f}_{c^{\prime},l^{\prime}}$ has a component $\parallel\bm{e}_{z}$ and $\bm{f}_{c,l}$ has a right-handed (RH) polarization component (a component $\parallel\bm{e}_{+}$). Turning now our attention to the mode $(c,-)$ given in Eqs.~\eqref{eq:VoigtModes}, we can rewrite the mode vector $\bm{f}_{c,-}$ as
\begin{equation}
\bm{f}_{c,-}=\frac{i}{\sqrt{2}}(1-\frac{\mathcal{F}[\omega_{c}]\mathcal{M}_{S}}{\varepsilon[\omega_{c}]})\bm{e}_{-}-\frac{i}{\sqrt{2}}(1+\frac{\mathcal{F}[\omega_{c}]\mathcal{M}_{S}}{\varepsilon[\omega_{c}]})\bm{e}_{+}.
\end{equation}
The mode $+$ is always linearly polarized along $\bm{e}_{z}$. Under the conditions of case (I), the mode $(c^{\prime},-)$ has a definite RH polarization, thus from our general considerations, it can not scatter into a higher frequency mode $(c,+)$. Otherwise, for case (II), the mode $(c,-)$ has a definite LH polarization and therefore it can not scatter to a lower frequency mode $(c^{\prime},+)$. This is a consequence of the conservation of polarization in the process: a Kittel magnon has a RH polarization, which yields the decomposition given by Eq.~\eqref{eq:DecompEpsPlus}. A photon with definite RH polarization can not annihilate a magnon with RH polarization generating a linearly polarized photon. These are \textquotedbl dispersion-tuned selection rules\textquotedbl{} that are valid at frequencies given by Eq.~\eqref{FreqSelectRules}. At such frequencies, a Brillouin light scattering experiment would exhibit a perfect transmission peak for an input light at the $-$ mode, without a peak at a lower (higher) frequency corresponding to the scattering into the $+$ mode. This is analogous to the selection rules fingerprints measured in scattering between whispering-gallery modes and magnon modes in magnetic spheres due to conservation of orbital angular momentum \cite{Sharma_2017_Light_scattering,Haigh_2018_Selection_rules,Osada_2018_Brillouin_light),Osada_2018_Orbital_angular}.

\begin{figure}
\centering{}\includegraphics[width=1.\columnwidth]{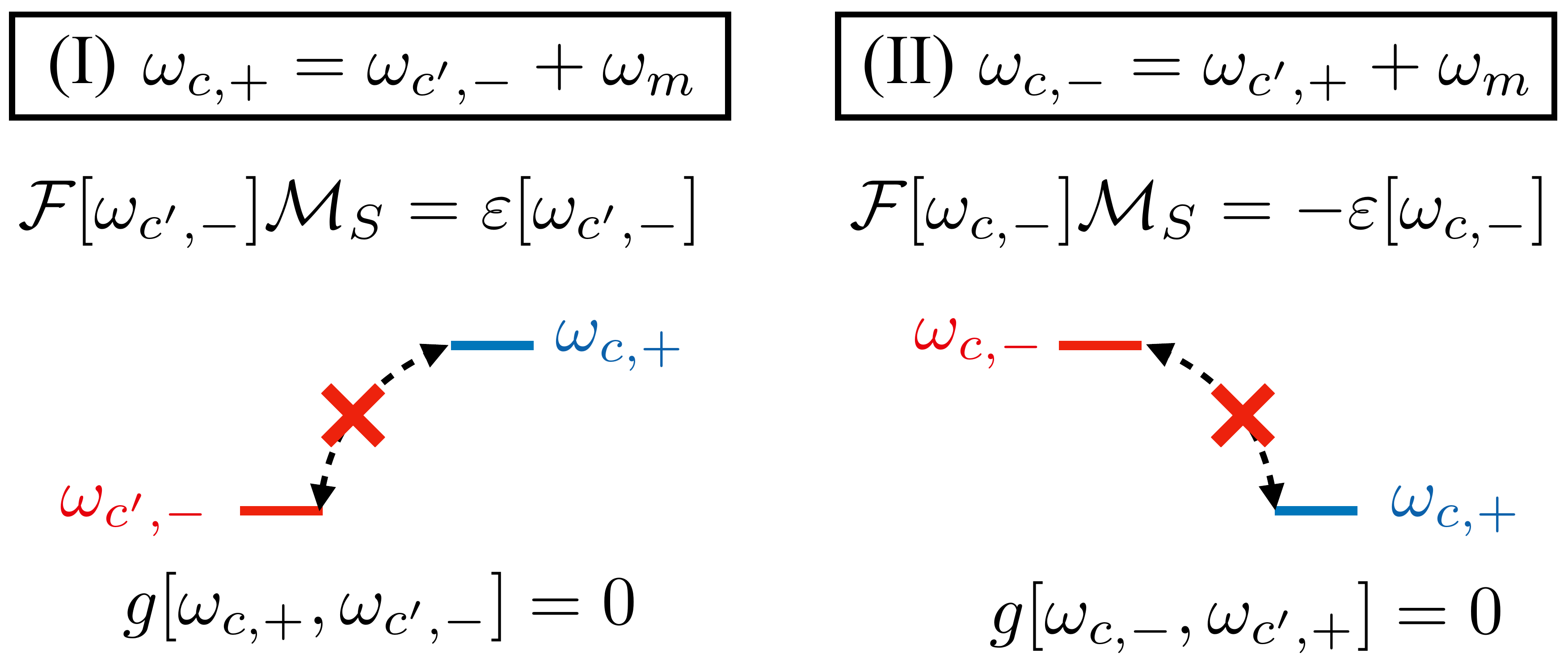}\caption{Schematic depicting the vanishing couplings between different polarized
Voigt modes due to dispersion.}
\label{SchematicVanishing}
\end{figure}

\subsubsection{Full Hamiltonian and Energy spectrum}

Considering the Hamiltonian describing the interaction between magnons and photons as derived in the previous sections, we can write the full Hamiltonian describing the photon-magnon system. We consider two optical modes and the uniform magnon mode, as was assumed for the derivations in the previous sections.

For the case in which the optical modes are degenerate the coupling term is given by Eq.~\eqref{eq:OptoMagCoupDege} and the full Hamiltonian reads
\begin{equation}
\begin{aligned}
\frac{\hat{H}_{{\rm D}}}{\hbar}&=\omega_{c}\left(\hat{a}_{+}^{\dagger}\hat{a}_{+}+\hat{a}_{-}^{\dagger}\hat{a}_{-}\right)+\omega_{m}\hat{m}^{\dagger}\hat{m}\\
&+ig_{{\rm Deg}}[\omega_{c}]\left(\hat{a}_{+}^{\dagger}\hat{a}_{-}-\hat{a}_{+}\hat{a}_{-}^{\dagger}\right)(\hat{m}^{\dagger}+\hat{m}).\label{eq:FullHamiltonianDegenerate}
\end{aligned}
\end{equation}
The first and last terms are obtained via the quantization procedure outlined above. The second term is obtained via the magnetic energy density with the Holstein-Primakoff transformation \cite{Holstein_1940_Field,Stancil_spin_2009,White_QuantumTheoryMag_2007}, and it takes into account exchange interactions, and an applied external bias field. An additional term $\propto\left(\hat{m}^{\dagger}\right)^{2}+\left(\hat{m}\right)^{2}$ can be included in the Hamiltonian in order to describe shape and crystalline anisotropy. We can find the energy spectrum of Eq.~\eqref{eq:FullHamiltonianDegenerate} by first performing a change of basis for the photon modes, $\hat{a}_{+}=i(\hat{a}_{1}+\hat{a}_{2})/\sqrt{2}$ and $\hat{a}_{-}=(\hat{a}_{1}-\hat{a}_{2})/\sqrt{2}$, and then, following \cite{Rabl_2011_Photon_Blockade,Nunnenkamp_2011_Single_Photon}, we perform a two-mode polaron transformation $\hat{H}\rightarrow\hat{U}^{\dagger}\hat{H}\hat{U}$, where $\hat{U}=e^{-\hat{S}}$ with 
\begin{equation}
\hat{S}=\frac{g_{{\rm Deg}}[\omega_{c}]}{\omega_{m}}(\hat{a}_{1}^{\dagger}\hat{a}_{1}-\hat{a}_{2}^{\dagger}\hat{a}_{2})(\hat{m}^{\dagger}-\hat{m}),
\end{equation}
such that the transformed Hamiltonian reads
\begin{equation}
\begin{aligned}
\frac{\hat{H}}{\hbar}  =\omega_m \hat{m}^\dagger \hat{m} &+\left(\omega_{c}-\frac{g_{{\rm Deg}}^{2}[\omega_{c}]}{\omega_{m}}\hat{a}_{1}^{\dagger}\hat{a}_{1}\right)\hat{a}_{1}^{\dagger}\hat{a}_{1}\\
&+\left(\omega_{c}-\frac{g_{{\rm Deg}}^{2}[\omega_{c}]}{\omega_{m}}\hat{a}_{2}^{\dagger}\hat{a}_{2}\right)\hat{a}_{2}^{\dagger}\hat{a}_{2} \\
 & +\frac{2g_{{\rm Deg}}^{2}[\omega_{c}]}{\omega_{m}}(\hat{a}_{1}^{\dagger}\hat{a}_{1})(\hat{a}_{2}^{\dagger}\hat{a}_{2}).\label{eq:HamilDiag}
\end{aligned}
\end{equation}
The eigenenergies are given in terms of the eigenvalues of the number operators $\hat{n}_{1,2}=\hat{a}_{1,2}^{\dagger}\hat{a}_{1,2}$ and $\hat{n}_{m}=\hat{m}^{\dagger}\hat{m}$ by
\begin{equation}
\begin{aligned}
\frac{E_{n_{1},n_{2},n_{m}}}{\hbar}=\omega_{m}n_{m}&+\omega_{c}(n_{1}+n_{2})+ \\
&-\frac{g_{{\rm Deg}}^{2}[\omega_{c}]}{\omega_{m}}(n_{1}-n_{2})^{2}\label{eq:EnergySpectrum}
\end{aligned}
\end{equation}
Each optical mode has a non-linear energy shift $\propto g_{{\rm Deg}}^{2}[\omega_{c}]n_{i}^{2}/\omega_{m}$, which becomes comparable with the single magnon transition energy if $g_{{\rm Deg}}^{2}[\omega_{c}]\sim\omega_{m}$, a regime attainable at $\omega_{{\rm ENZ}}$ provided that the zero point magnetization fluctuations are large, which depends on the cavity volume. From Eq.~\eqref{eq:EZCoupling}, since $\mathcal{M}_{{\rm ZPF}}/\mathcal{M}_{S}=\left(\gamma\hbar/2V\mathcal{M}_{S}\right)^{1/2}$, $g_{{\rm Deg}}[\omega_{{\rm ENZ}}]\sim\omega_{m}$ can be achieved with cavity volumes 
\begin{equation}
V\sim\frac{\omega_{{\rm ENZ}}^{2}}{\omega_{m}^{2}}\frac{\hbar\gamma}{2\mathcal{M}_{S}}.
\end{equation}
The saturation magnetization depends on the material, while the magnon frequency $\omega_{m}$ is defined by an external bias magnetic field $H_{{\rm Ex}}$ via $\omega_{m}=\mu_{0}\gamma H_{{\rm Ex}}$. The ENZ frequency depends on the specific dispersion model. Taking as an example the saturation magnetization of Yttrium-Iron Garnet (YIG), $\mathcal{M}_{S}=196\,{\rm kA/m}$ \cite{Stancil_spin_2009} we can estimate $V\sim4.7\times10^{-12}\omega_{{\rm ENZ}}^{2}/\omega_{m}^{2}\,{\rm \mu m^{3}}$. Thus, in the typical case where $\omega_{m}\sim2\pi\times10\:{\rm GHz}$, $g_{{\rm Deg}}[\omega_{{\rm ENZ}}]\sim\omega_{m}$ can be achieved in a cavity with a volume $\sim\mu{\rm m}^{3}$ provided that the ENZ frequency is $\omega_{{\rm ENZ}}\sim 2 \pi \times 1400\:{\rm THz}$, which is the case considered in Ref. \cite{Bittencourt_2021_ENZ_Lett} for the dispersion model of Section \ref{sec:Lorentz-dispersion-for}.

The fingerprints of the non-linear energy dispersion can be probed via the power spectrum of one optical mode under an external coherent drive as we report in \cite{Bittencourt_2021_ENZ_Lett}. For that, the magnon sidebands in the power spectrum generated by the energy dispersion given in Eq.~\eqref{eq:EnergySpectrum} need to be resolved, which requires the single-magnon strong coupling regime where the coupling is larger than the magnon ($\kappa_{m}$) and photon decay rates ($\kappa$). The magnon decay rate depends on the Gilbert damping and it is typically $\sim{\rm MHz}$. The photon decay rate depends on the intrinsic material losses and on cavity design (radiative decay). The optical decay in optimized state-of-the-art systems are $\sim{\rm GHz}$. Since $\kappa<\omega_{m}$, the most stringent requirement in terms of volume and ENZ frequency is the one for $g[\omega_{{\rm ENZ}}]\sim\omega_{m}$. For YIG inspired parameters, the estimate of $\omega_{\rm{ENZ}}$ presented in the previous paragraph and the requirements discussed above correspond to high-frequency photons, which would induce transitions between the bands of the material \cite{Li_2021_First_principles}, introducing an additional dissipation channel. Nevertheless, experiments with Cerium-substituted YIG films \cite{Onbasli_2016_optical_and_MO}, indicates a large value of the Faraday effect per unit length at such frequencies. Given those considerations, prospect implementations of such system would incorporate the magneto-optical effects of magnetic dielectrics in layered structures exhibiting an effective ENZ behavior \cite{davoyan_theory_of_2013,giron_giant_enhancement_2017,Almpanis_controllingTMOKE_2020,Moncada_2020_UniaxialENZ} or in wave-guides \cite{davoyan_optical_isolation_2013,davoyan_nonreciprocal_emission_2019}. Since our theory is valid despite the origin of dispersion, it can also be applied to such structured media.

The strong single particle coupling in this system yields the non-linear energy spectrum of  Eq.~\eqref{eq:EnergySpectrum}, which in turn generates photon blockade in optomagnonical cavities. The cavity frequency is shifted by the number of photons, thus after a first photon is added to the cavity, by an external source,  the non-linear energy shift prevents a second and multiple photons to enter the cavity \cite{Rabl_2011_Photon_Blockade}. Such effect can be probed by the measurement of the cavity output field statistics which, in this case, would be anti-bunched. Another possibility for probing the strong single-magnon coupling is via magnon-induced transparency. In this case, a control drive with frequency $\omega_D$ changes the transmission of a second weaker probe drive with frequency $\omega_p$. Induced transparency was described for single-particle weak coupling cavity optomagnonics \cite{Liu_2016_Optomagnonics}, in which the control is tuned at the first red magnon sideband $\omega_D = \omega_c - \omega_m$, generating an interference with a probe input at the cavity frequency, such that the latter is perfectly transmitted. This can be measured via a transmission spectrum as a function of the probe frequency which has a transparency peak at the cavity frequency. When the system exhibits a strong single-magnon coupling rate, such standard induced transparency would be modified due to the non-linear shift in the transitions, yielding a modification in the transparency peak at the cavity frequency \cite{Kronwald_2013_Optomechanically_Induced}. Furthermore, an induced transparency signal would be obtained at the second magnon sideband, that is, for $\omega_p=\omega_c - 2 \omega_m$, for a control drive at the cavity frequency $\omega_D=\omega_c$ \cite{Kronwald_2013_Optomechanically_Induced}. Such signal is a consequence of two-photon transitions driven by the non-linear coupling term in Eq.~\eqref{eq:FullHamiltonianDegenerate}, and would be an unambiguous fingerprint of single-magnon strong coupling in an ENZ optomagnonical cavity. Finally, the strong magnon-photon coupling regime can be harnessed for the deterministic optical generation of non-classical magnon states \cite{Nunnenkamp_2011_Single_Photon}.

For the case of non-degenerate optical modes, it is not possible to find the energy spectrum analytically. Nevertheless, if the frequency difference between the optical modes $\omega_{-}-\omega_{+}$ is comparable to the magnon frequency $\omega_{m}$ and the coupling $g\ll\omega_{m}$, an effective Hamiltonian can be obtained by performing a Schrieffer-Wolf like transformation \cite{Ludwig_2012_Enhanced_quantum}. In this case, the effective interaction between photons and magnons can be used to realize a quantum non-demolition measurement of the magnon number \cite{Ludwig_2012_Enhanced_quantum}.

\section{Lorentz dispersion model for a Faraday-active medium\label{sec:Lorentz-dispersion-for}}

We now describe the features of the frequency-dependent optomagnonic coupling in a specific dispersion model based on the dispersion obtained in paramagnetic and ferromagnetic dielectrics. The frequency dependence of the permittivity tensor is obtained following Ref.~\cite{Crossley_faraday_1969}, which yields a permittivity tensor linear in the magnetization that takes into account the Faraday effect. This framework does not includes the Cotton-Mouton effect.

To derive the dispersion of the optical properties of a magnetic dielectric, we consider the linear response of the medium to an electromagnetic field defined via the polarization
\begin{equation}
P_{l}(\bm{r},t)=\langle\hat{P}_l(\bm{r},t)\rangle_{{\rm Th}}=\varepsilon_{0}\int_{-\infty}^{t}dt^{\prime}\chi_{lk}(t-t^{\prime})E_{k}(\bm{r},t^{\prime}),\label{eq:RespFunc01-1-1}
\end{equation}
where $l,k=(x,y,z)$, the average value is over the thermal state of the system at $t=-\infty$, and $\overleftrightarrow{\chi}[\omega]$ is the susceptibility tensor. The permittivity tensor is given by $\overleftrightarrow{\varepsilon}[\omega]=1+\overleftrightarrow{\chi}[\omega]$. At optical frequencies, the only transitions leading to dispersion in our model are electric dipole transitions. Magnetic dipole transitions, that would generate dispersion in the magnetic permeability, are far off-resonant and we can be safely disregarded. At some intermediate frequency it is possible that dispersion due to both magnetic and electric dipole transitions might be present, a framework which we do not study here. This case can be particularly important for structured media with effective properties mimicking those of continuum bulk media and can be relevant for describing optomagnonics in near-zero index (NZI) media, in which both the permittivity and the permeability vanish \cite{liberal_near_zero_2017,kinsey_near_zero_index_2019}.

The susceptibility tensor is a response function that can be calculated from the eigenstates of the system's Hamiltonian. We assume that inside the infinitesimal volume surrounding a point $\bm{r}$, the electric field does not vary appreciably and that the volume contains a number $N$ of magnetic ions. We then compute the contribution of each individual ion to the response of the material. For a single ion, the interaction with light is given by the usual dipole interaction in the long wavelength approximation
\begin{equation}
\hat{H}=\hat{H}_{{\rm Ion}}+\bm{\hat{V}}\cdot\bm{E}(\bm{r},t),
\end{equation}
were $\hat{\bm{V}}=-e\bm{r}$ is the dipole operator and $\hat{H}_{{\rm Ion}}$ includes all terms defining the internal ionic levels \cite{Fleury_scatteringoflight_1968,Shen_faraday_rot_1964,Gall_spinphoton_1971}:
\begin{equation}
\hat{H}_{{\rm Ion}}=\hat{H}_{0}+\hat{H}_{1},
\end{equation}
where $\hat{H}_{{\rm 0}}$ defines a set of unsplit ground states denoted by $\left\{ \vert g\rangle\right\} $ and a set of excited states $\left\{ \vert e\rangle\right\} $. The perturbation term $\hat{H}_{1}$ splits the ground state degeneracy, it can contain spin-orbit coupling, crystalline fields, the Zeeman interaction (with an external bias magnetic field) and exchange fields. The crystalline anisotropy, Zeeman and exchange interactions break the degeneracy of the ground state manifold, while the spin-orbit and the Zeeman interactions break the degeneracy of the excited states. Spin-orbit coupling is particularly important for the Faraday effect: it mixes excited states with different spin components, allowing transitions between ground states with different magnetic moments, e.g., from $m_{z}=-1/2$ to $m_{z}=+1/2$. 

We assume transitions between a group of ground states and a group of excited states. From the Kubo formula, the component $lk$ ($l,k=x,y,z$) of the susceptibility is given by $\chi_{lk}(t-t^{\prime})=\frac{i}{\hbar}\Theta(t-t^{\prime})\langle\left[\hat{\tilde{V}}_{l},\hat{\tilde{V}}_{k}\right]\rangle_{{\rm Th}}$ (the tildes indicates operators in the interaction picture). Assuming that only the lower energy levels of the orbitals are occupied, one gets 
\begin{equation}
\begin{aligned}
\chi_{kl}[\omega]=\sum_{g,e}\rho_{g}X(\omega,\omega_{eg})\Big[&\frac{\omega_{eg}}{\omega}{\rm Re}\left(\langle g\vert\hat{V}_{k}\vert e\rangle\langle e\vert\hat{V}_{l}\vert g\rangle\right)\\
+&i{\rm Im}\left(\langle g\vert\hat{V}_{k}\vert e\rangle\langle e\vert\hat{V}_{l}\vert g\rangle\right)\Big],\label{eq:Suscept-1-1}
\end{aligned}
\end{equation}
with $\rho_{g}$ the Boltzmann factor giving the population of a given ground state $\vert g\rangle$, and $X(\omega,\omega_{eg})$ a complex shape factor for the transition between the relevant states given by
\begin{equation}
X(\omega,\omega_{e_{j}g_{j}})=\frac{2\omega}{\hbar(\omega_{e_{j}g_{j}}^{2}-\omega^{2}-2i\omega\eta_{e_{j}g_{j}})}.
\end{equation}

The response function depends on the dipole matrix transition elements $\langle g\vert\hat{V}_{k}\vert e\rangle$, which in turn are defined by the perturbation terms included in $\hat{H}_{{\rm 1}}$. First order perturbation terms, such as the one arising from spin-orbit coupling, contributes to inelastic scattering and generates the Faraday effect \cite{Crossley_faraday_1969,Shen_faraday_rot_1964,Fleury_scatteringoflight_1968,Gall_spinphoton_1971}. Second order scattering processes, for example arising from exchange interaction, contributes to the Cotton-Mouton effect.

To obtain the dispersion properties of the permittivity tensor, we first write the transition frequencies between an excited and a ground state as
\begin{equation}
\omega_{eg}=\omega_{0}+\frac{\Delta E_{e}-\Delta E_{g}}{\hbar},
\end{equation}
where $\hbar\omega_{0}$ is the energy difference between the energy levels without the perturbations and $\Delta E_{e(g)}^{\prime}=E_{e(g)}^{\prime}-E_{e(g)}$ are the energy differences between perturbed and unperturbed levels. We also assume that the magnetization is saturated along the $\bm{e}_z$ direction. Assuming that the perturbations responsible for breaking the level degeneracies are small, we take $\Delta E_{e},\Delta E_{g}\ll\vert\hbar\omega-\hbar\omega_{0}+i\omega\eta_{eg}\vert$, and since ${\rm Im}\left[\langle g\vert\hat{V}_{+}\vert e\rangle\langle e\vert\hat{V}_{+}\vert g\rangle\right]=0$, the components of the susceptibility tensor are given by
\begin{equation}
\begin{aligned}
\chi_{jj}[\omega] & =N\frac{2\omega_{0}\mathcal{A}_{j}}{\hbar\left(\omega_{0}^{2}-\omega^{2}+i2\omega\eta\right)},\\
\chi_{xy}[\omega] & =N\sum_{e,g}\frac{i\rho_{g}\omega}{\hbar\left(\omega_{0}^{2}-\omega^{2}+i2\omega\eta\right)}\\
&\quad \quad \times \left[1-2\omega_{0}\frac{\Delta E_{e}-\Delta E_{g}}{\hbar\left(\omega_{0}^{2}-\omega^{2}+i2\omega\eta\right)}\right]\left(f_{-}^{eg}-f_{+}^{eg}\right),
\end{aligned}
\end{equation}
where $\mathcal{A}_{j}=\sum_{e,g}\rho_{g}\vert\langle g\vert\hat{V}_{j}\vert e\rangle\vert^{2}$ and $f_{\pm}=\frac{1}{2}\langle g\vert\hat{V}_{\pm}\vert e\rangle\langle e\vert\hat{V}_{\mp}\vert g\rangle=\frac{1}{2}\vert\langle g\vert\hat{V}_{\pm}\vert e\rangle\vert^{2}$. For simplicity we take the diagonal components of the susceptibility as
\begin{equation}
\chi_{jj}[\omega]=\frac{2\omega_{0}N\mathcal{A}}{\hbar\left(\omega_{0}^{2}-\omega^{2}+i2\omega\eta\right)}.
\end{equation}

The off diagonal component $\chi_{xy}$ can be computed by first writing
\begin{equation}
\chi_{xy}[\omega]  =\chi_{xy}^{(1)}[\omega]+\chi_{xy}^{(2)}[\omega]+\chi_{xy}^{(3)}[\omega],
\label{eq:SuceptFull}
\end{equation}
where
\begin{equation}
\begin{aligned}
\chi_{xy}^{(1)}[\omega] & =\frac{iN\omega \sum_{e,g}\rho_{g}\left(f_{-}^{eg}-f_{+}^{eg}\right)}{\hbar\left(\omega_{0}^{2}-\omega^{2}+i2\omega\eta\right)},\\
\chi_{xy}^{(2)}[\omega] & =\frac{i 2 \omega_{0} N \omega \sum_{g,e}\rho_{g} \Delta E_{g}\left(f_{-}^{eg}-f_{+}^{eg}\right)}{\hbar\left(\omega_{0}^{2}-\omega^{2}+i2\omega\eta\right)^{2}},\\
\chi_{xy}^{(3)}[\omega] & =-\frac{i 2 \omega_{0} N \omega \sum_{g,e}\rho_{g}\Delta E_{e}\left(f_{-}^{eg}-f_{+}^{eg}\right)}{\hbar\left(\omega_{0}^{2}-\omega^{2}+i2\omega\eta\right)^{2}}.
\end{aligned}
\end{equation}
We define $\hat{P}_{e}=\sum_{_{e}}\vert e\rangle\langle e\vert$, such that $\sum_{e}f_{\pm}^{eg}=\langle g\vert\hat{V}_{\pm}\hat{P}_{e}\hat{V}_{\mp}\vert g\rangle$. Furthermore, the energy corrections are given by first order perturbation theory $\Delta E_{e(g)}=\langle e(g)\vert\hat{H}_{1}\vert e(g)\rangle$ such that $\sum_{e}\Delta E_{e}f_{\pm}^{eg}=\langle g\vert\hat{V}_{\pm}\hat{P}_{e}\hat{H}_{I}\hat{P}_{e}\hat{V}_{\mp}\vert g\rangle$. The non-vanishing terms of the susceptibility depend on specific characteristics of the ground and excited states defined by the terms included in $\hat{H}_{0}$, and the calculation of $\langle g\vert\hat{V}_{\pm}\hat{P}_{e}\hat{V}_{\mp}\vert g\rangle$ and $\langle g\vert\hat{V}_{\pm}\hat{P}_{e}\hat{H}_{I}\hat{P}_{e}\hat{V}_{\mp}\vert g\rangle$ follows the procedure outlined in Ref.~\cite{Crossley_faraday_1969}. It can be show that Eq.~\eqref{eq:SuceptFull} is proportional to the ground state magnetic moment $M_{z}$, such that the off diagonal component of the susceptibility reads
\begin{equation}
\begin{aligned}
\chi_{xy}[\omega]&=i\mathcal{F}[\omega]\left(NM_{z}\right) \\
&=i\frac{2\omega\left(NM_{z}\right)}{\hbar\left(\omega_{0}^{2}-\omega^{2}+i2\omega\eta\right)}\left[A_{2}+\frac{A_{3}\omega_{0}}{\hbar\left(\omega_{0}^{2}-\omega^{2}+i2\omega\eta\right)}\right],
\end{aligned}
\end{equation}
where
\begin{equation}
\mathcal{F}[\omega]=\frac{2\omega}{\left(\omega_{0}^{2}-\omega^{2}+i2\omega\eta\right)}\left[A_{2}+\frac{A_{3}\omega_{0}}{\left(\omega_{0}^{2}-\omega^{2}+i2\omega\eta\right)}\right].\label{eq:eqF}
\end{equation}
The calculations can be generalized to include the other components of the susceptibility tensor which give the permittivity tensor in the form of Eq.~\eqref{eq:PermitividadeEletrica}. In this model for the dispersion of a Faraday-active dielectric, $A_{2}$ depends on the deviations of the g-factor of the ion from $2.002$ and is only relevant for ground states with both non-zero orbital angular momentum and splitting due to crystalline anisotropy. For ions with ground states with zero orbital angular momentum, $A_{2}=0$ and the only term contributing to the Faraday effect is the one proportional to $A_{3}$, which is given in terms of the spin-orbit coupling constant for the excited state manifold. The saturation magnetization is $NM_{z}=\mathcal{M}_{S}$. The formula is also valid for paramagnets and can be generalized to include contributions from different types of ions, which extends the description to ferrimagnets. We consider here that the medium in which the electromagnetic waves propagate is described by the above dispersion model.

In the limit $\omega\ll\omega_{0}$, $\varepsilon_{jj}[0]=1+\chi_{jj}[0]=\bar{\varepsilon}$ is the electrostatic permittivity \cite{Landau_electrodynamics_2009}. Assuming equal values of the permittivity for all diagonal components we obtain
\begin{equation}
\mathcal{A}=\frac{\hbar\omega_{0}}{2N}\left(\bar{\varepsilon}-1\right)
\end{equation}
and therefore
\begin{equation}
\chi_{jj}[\omega]=\frac{\omega_{0}^{2}\left(\bar{\varepsilon}-1\right)}{\left(\omega_{0}^{2}-\omega^{2}+i2\omega\eta\right)}.
\end{equation}
The off-diagonal terms define the Faraday rotation angle per length as
\begin{equation}
\Phi[\omega]=\frac{\omega}{2c}(n_{-}[\omega]-n_{+}[\omega])=i\frac{\omega}{2c}\frac{\chi_{xy}[\omega]-\chi_{yx}[\omega]}{2\bar{n}[\omega]},\label{eq:FarRotAng}
\end{equation}
where $n_{\pm}[\omega]=\sqrt{\varepsilon[\omega]\pm\mathcal{F}[\omega]\mathcal{M}_{S}}$ are the refractive indexes for positive and negative circular polarizations and $\bar{n}[\omega]=(n_{-}[\omega]+n_{+}[\omega])/2$ is the average refractive index of the medium. At frequencies $\omega\ll\omega_{0}$, $\bar{n}[\omega]\simeq n[\omega]=\sqrt{1+\chi[\omega]}=\sqrt{\varepsilon[\omega]}$. We can write for a magnet with magnetization saturated along $\bm{e}_{z}$

\begin{equation}
\Phi[\omega]=-\frac{\omega}{2c}\frac{\mathcal{F}[\omega]}{\sqrt{\varepsilon[\omega]}}\mathcal{M}_{s}.\label{eq:FaradayPerLengthUsual}
\end{equation}
The Verdet constant is given by $\mathcal{V}[\omega]=-\frac{\omega}{2c}\frac{\mathcal{F}[\omega]}{\sqrt{\varepsilon[\omega]}}$. We can then write compactly
\begin{align}
\chi_{xy}[\omega] & =i\mathcal{F}[\omega]\mathcal{M}_{z},\label{eq:Suscept2}\\
\varepsilon[\omega] & =1+\frac{\omega_{0}^{2}\left(\bar{\varepsilon}-1\right)}{\left(\omega_{0}^{2}-\omega^{2}+i2\omega\eta\right)}.\nonumber 
\end{align}

Before proceeding, we introduce the adimensional parameters $\tilde{\eta}=\eta/\omega_{0}$, $\tilde{A}_{2}=A_{2}\mathcal{M}_{S}/\omega_{0}$ and $\tilde{A}_{3}=A_{3}\mathcal{M}_{S}/\omega_{0}^{2}$. The weakness of optomagnonic effects implies that $\tilde{A}_{2,3}\ll1$. For the parameters corresponding to the $500\:{\rm nm}$ transition of octahedral oriented ${\rm Fe}^{3+}$ ions in YIG, we have $\tilde{A}_{3}\sim10^{-4}$ \cite{Crossley_faraday_1969,scott_absorptionspectra_1974,Scott_magneticcircular_1975}. This particular case corresponds to a ground state with no orbital angular momentum, and thus $\tilde{A}_{2}=0$. Here, we study a more general framework and describe features from both $\tilde{A}_{2}$ and $\tilde{A}_{3}$ terms. Furthermore, even though the above model is based on ionic transitions between a set of ground and excited states, a similar dependence of the permittivity on the frequency describes the properties of other systems such as magnetized plasmas \cite{Shen_analogof_2019} and metal-dielectric-gyroelectric heterostructures \cite{Tsakmakidis_2017_breaking_lorentz}.

\begin{figure*}
\centering{}\includegraphics[width=2.\columnwidth]{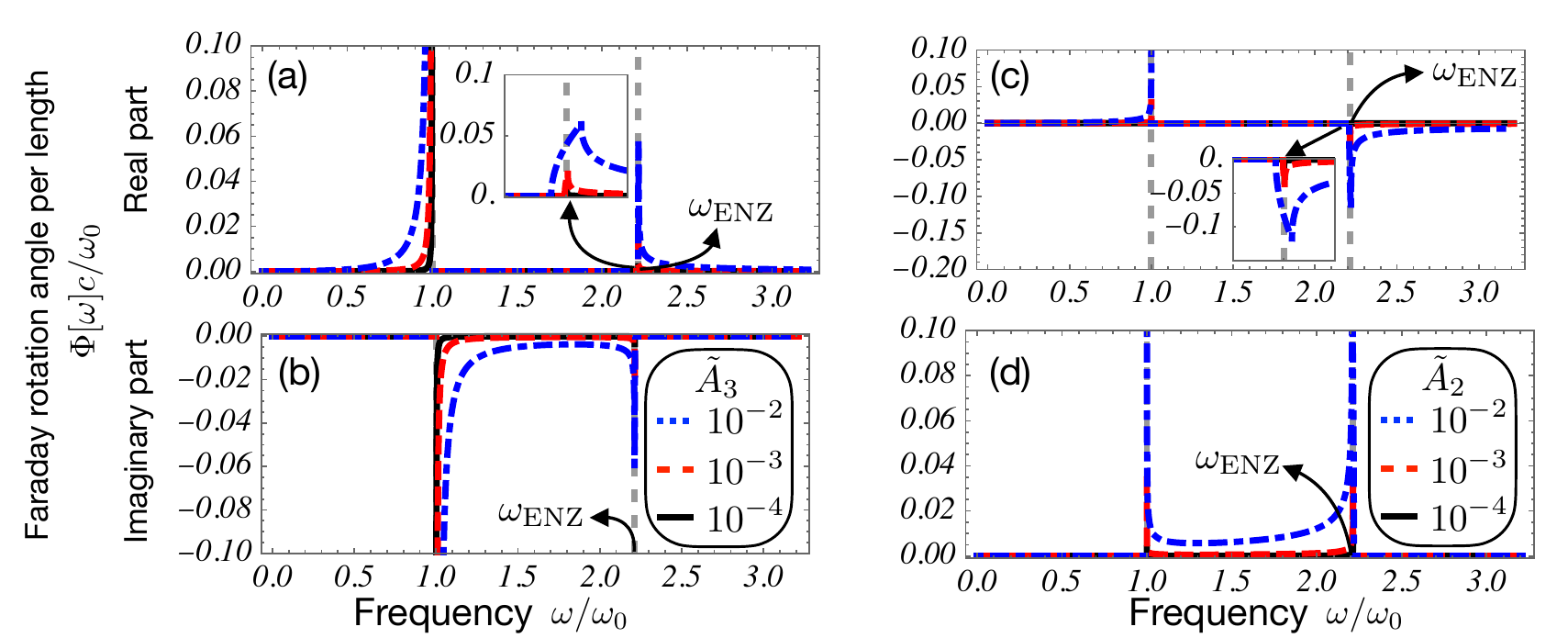}\caption{Faraday rotation angle per unit length (see Eq.~\ref{eq:FarRotAng}) in units of $\omega_{0}/c$ as a function of the frequency in units of $\omega_{0}$ for (a,b) $\tilde{A}_{2}=0$ and (c,d) $\tilde{A}_{3}=0$. The insets show a zoom-in around the ENZ frequency $\omega_{{\rm ENZ}}$. For both cases the Faraday rotation angle is enhanced in this frequency region, but does not diverge. For these plots we adopted $\bar{\varepsilon}=4.9$.}
\label{Fig02:Permittivity}
\end{figure*}

The diagonal part of $\varepsilon[\omega]$ follows the usual Lorentz-model behavior and vanishes at $\omega=\omega_{{\rm ENZ}}=\omega_{0}\sqrt{\bar{\varepsilon}}(1-\eta^{2}/2\omega_{0}^{2})$, which we call the ENZ frequency. In what follows we consider the case $\tilde{\eta}\ll1$, such that we can neglect its effect as long as the frequency range is far from the resonance frequency $\omega_{0}$. In this case $\omega_{{\rm ENZ}}=\omega_{0}\sqrt{\bar{\varepsilon}}$. Similarly to $\varepsilon[\omega]$, whereas the Faraday rotation angle also diverges at $\omega_{0}$ for $\eta\rightarrow0$, it is finite around $\omega_{{\rm ENZ}}$. This is shown in Fig.~\ref{Fig02:Permittivity} where we plot the Faraday rotation angle for $\tilde{A}_{2}=0$ (Fig.~\ref{Fig02:Permittivity}a,b) and for $\tilde{A}_{3}=0$ (Fig~\ref{Fig02:Permittivity}c,d). The former case corresponds to the framework of Ref.~\cite{Bittencourt_2021_ENZ_Lett}. 

\subsection{Characterization of the plane wave modes for the Lorentz dispersion model \label{subsec:Optomagnonic-coupling-for}}

Before evaluating the optomagnonic coupling for the Lorentz-like dispersion model, we discuss the propagation characteristics of the plane-wave modes obtained in \ref{subsec:Plane-wave-like-modes:} for both Faraday and Voigt configurations.

\begin{figure}
\centering{}\includegraphics[width=1.\columnwidth]{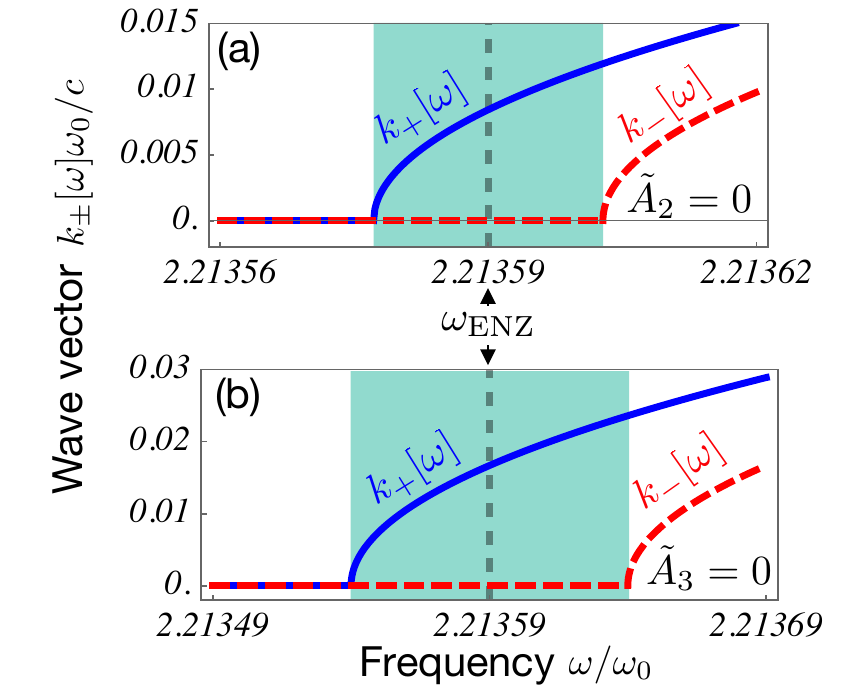}\caption{Wave vectors (real part) for EM plane waves propagating in the Faraday configuration for the dispersion model of Eq.~\eqref{eq:Suscept2} as a function of the frequency (in units of $\omega_{0}$) for a frequency range close to the ENZ frequency. (a) for $\tilde{A}_{2}=0$ and $\tilde{A}_{3}=10^{-4}$ and (b) for $\tilde{A}_{3}=0$ and $\tilde{A}_{2}=10^{-4}$. The shaded regions indicate the frequency range in which only one polarization propagates. We have adopted $\bar{\varepsilon}=4.9$. Frequency in units of the ionic transition frequency $\omega_0$, wave vectors in units of $c/\omega_0$.}
\label{FigurekFaraday}
\end{figure}

\begin{figure}
\begin{centering}
\includegraphics[width=1.\columnwidth]{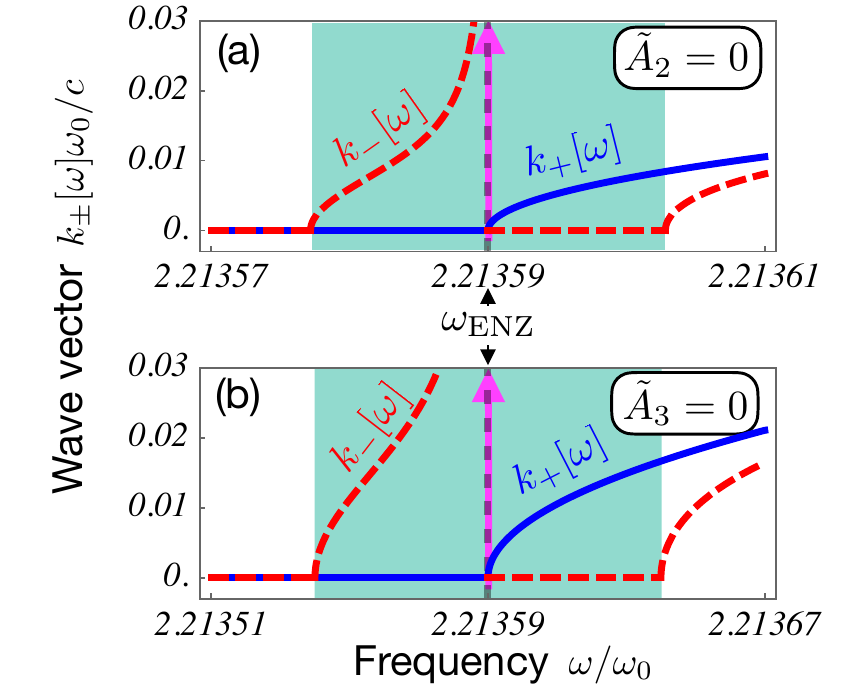}\caption{Wave vectors (real part) for EM plane waves propagating in the Voigt configuration for the dispersion model of Eq.~\eqref{eq:Suscept2} as a function of the frequency (in units of $\omega_{0}$) for a frequency range close to the ENZ frequency. (a) for $\tilde{A}_{2}=0$ and $\tilde{A}_{3}=10^{-4}$ and (b) for $\tilde{A}_{3}=0$ and $\tilde{A}_{2}=10^{-4}$. The shaded regions indicate the frequency range in which only one polarization propagates. We adopted $\bar{\varepsilon}=4.9$. Frequency in units of the ionic transition frequency $\omega_0$, wave vectors in units of $c/\omega_0$.}
\label{FigurekVoigt}
\par\end{centering}
\end{figure}

Figure \ref{FigurekFaraday} shows the wave vector for electromagnetic waves propagating in the Faraday configuration for (a) $\tilde{A}_{2}=0$ and (b) $\tilde{A}_{3}=0$. We notice a frequency range (shaded region in the figure) in which only one of the modes has a real wave vector, while the other exhibits a pure imaginary wave vector. In this range, only the $(+)$ mode propagates, while the $(-)$ mode is evanescent. This is the basis for design an magneto-optical based isolator in the Faraday configuration \cite{davoyan_optical_isolation_2013}. A similar effect is also present for EM waves in the Voigt configuration, as depicted in Fig.
~\ref{FigurekVoigt}. The wave vector of the Voigt mode $(-)$ diverges at $\omega_{{\rm ENZ}}$ due to the term $\propto\mathcal{F}^{2}[\omega]/\varepsilon[\omega]$ (c.f. Eq.~(\ref{eq:VoigtK})) for both cases depicted in Fig. \ref{FigurekVoigt}. From Eq.~\eqref{eq:VoigtModes}, we also notice that at $\omega_{{\rm ENZ}}$ the polarization of this mode is $\bm{f}_{-}\propto\bm{e}_{x}\parallel\bm{k}_{-}$, implying that the mode describes perfectly longitudinal waves. This is due to the frequency-dependent circular magnetic birefringence. This is in contrast with the Faraday case, which exhibit finite wave vectors for both modes at the ENZ frequency and always has (discarding terms $\propto\mathcal{M}_{x,y}$) perfectly circular polarization. Furthermore, in the Voigt configuration there is also a region in which one of the modes propagates while the other is evanescent, yielding, similar to the Faraday case, to opical isolation. 

\subsection{Optomagnonic coupling for the Lorentz dispersion: Degenerate case}

\begin{figure}
\centering{}\includegraphics[width=1.\columnwidth]{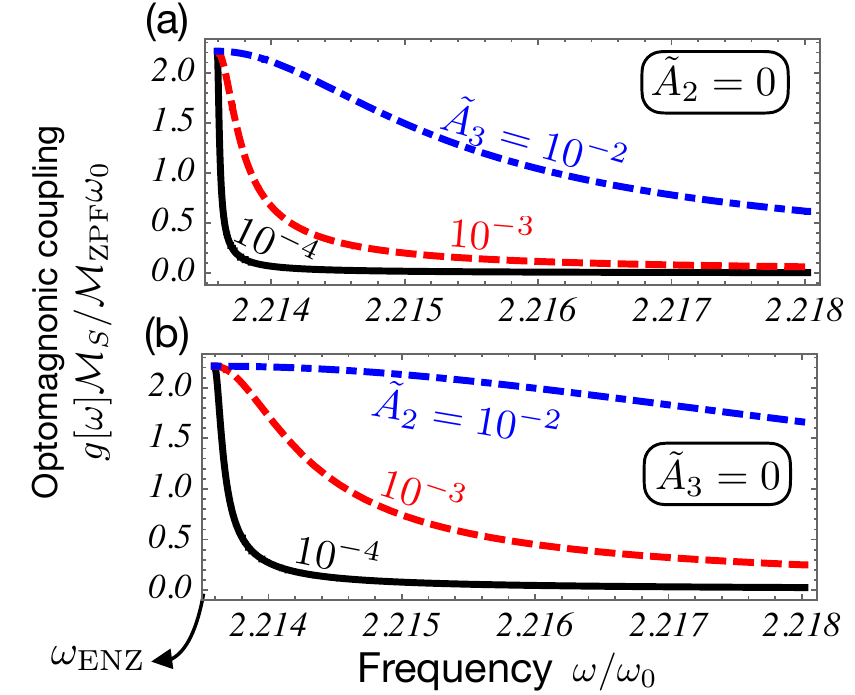}\caption{Optomagnonic coupling for degenerate modes as a function of the frequency, for frequencies close to the ENZ point. (a) for $\tilde{A}_{2}=0$ and (b) for $\tilde{A}_{3}=0$, see Eqs.~\eqref{eq:Suscept2} and \eqref{eq:eqF}. For these plots we adopted $\bar{\varepsilon}=4.9$. Frequency in units of the ionic transition frequency $\omega_0$,  and the coupling is given in units of $\mathcal{M}_{{\rm ZPF}}\omega_{0}/\mathcal{M}_{S}$.}
\label{Fig:CouplingDegenerate}
\end{figure}

Given the dispersion model we can study the behavior of the optomagnonic coupling in the Voigt modes as a function of the frequency. The coupling for degenerate modes, see Eq.~\eqref{eq:CoupVoiu}, is shown in Fig.~\ref{Fig:CouplingDegenerate} using the expressions for $\varepsilon$ and $\mathcal{F}$ given in Eq.~\eqref{eq:Suscept2}, and for several values of $\tilde{A}_{2}$ and $\tilde{A}_{3}$. At $\omega=\omega_{{\rm ENZ}}$, as shown in \ref{Fig:CouplingDegenerate}, the coupling is $\mathcal{M}_{{\rm ZPF}}\omega_{{\rm ENZ}}/\mathcal{M_{S}}$. The prefactor for the adopted values is $\mathcal{M}_{{\rm ZPF}}/\mathcal{M}_{S}\sim 6.8\times10^{-15}/\sqrt{V}$. Smaller magnetic volumes yield stronger couplings, as we discussed in Subsection \ref{subsec:Optomagnonic-coupling-Hamiltonia}. In particular, for illustrative parameters of YIG, $\tilde{A}_{2}\sim0$, $\tilde{A}_{3}\sim10^{-4}$, $\bar{\varepsilon}=4.9$, and for a volume of $\sim(\mu{\rm m})^{3}$ it can be shown that the coupling is $g \sim 2 \pi \times 10$ GHz, comparable to the magnon frequency as reported in Ref.~\cite{Bittencourt_2021_ENZ_Lett}. Even though one of the wave vectors for the Voigt modes exhibits a diverging behavior at $\omega_{{\rm ENZ}}$, the coupling remains finite. 

For frequencies $\omega\ll\omega_{0}$, $\partial_{\omega_{c}}(\omega_{c}\varepsilon[\omega_{c}])\sim\bar{\varepsilon}$ and 
\begin{equation}
\begin{aligned}
\frac{\mathcal{F}[\omega_{c}]}{\varepsilon[\omega_{c}]}\partial_{\omega_{c}}\left(\omega_{c}\varepsilon[\omega_{c}]\right)-\partial_{\omega_{c}}\left(\omega_{c}\mathcal{F}[\omega_{c}]\right) &\sim\\
&\mathcal{F}[\omega_{c}]-\partial_{\omega_{c}}\left(\omega_{c}\mathcal{F}[\omega_{c}]\right)\\
&=-\omega_{c}\partial_{\omega_{c}}\left(\mathcal{F}[\omega_{c}]\right).
\end{aligned}
\end{equation}
In this limit and for the Lorentz-like dispersion model, $\mathcal{F}[\omega_{c}]\sim\omega_{c}F$ (with $F$ a constant), thus $\omega_{c}\partial_{\omega_{c}}\left(\mathcal{F}[\omega_{c}]\right)\sim\mathcal{F}[\omega_{c}]$. Therefore $C_{-}\sim C_{+}=1/\sqrt{\varepsilon_{0}\bar{\varepsilon}}$, and the obtained coupling in terms of the Faraday rotation angle per length reads
\begin{equation}
G\sim\frac{c\theta_{F}}{\sqrt{\bar{\varepsilon}}}\,,
\end{equation}
consistent with the expression used in the optomagnonics bibliography in the non-dispersive regime \cite{Liu_2016_Optomagnonics,Kusminskiy_2016_Coupled}. 

\subsection{Optomagnonic coupling for the Lorentz dispersion: Non-degenerate case}

\begin{figure*}
\begin{centering}
\includegraphics[width=2.\columnwidth]{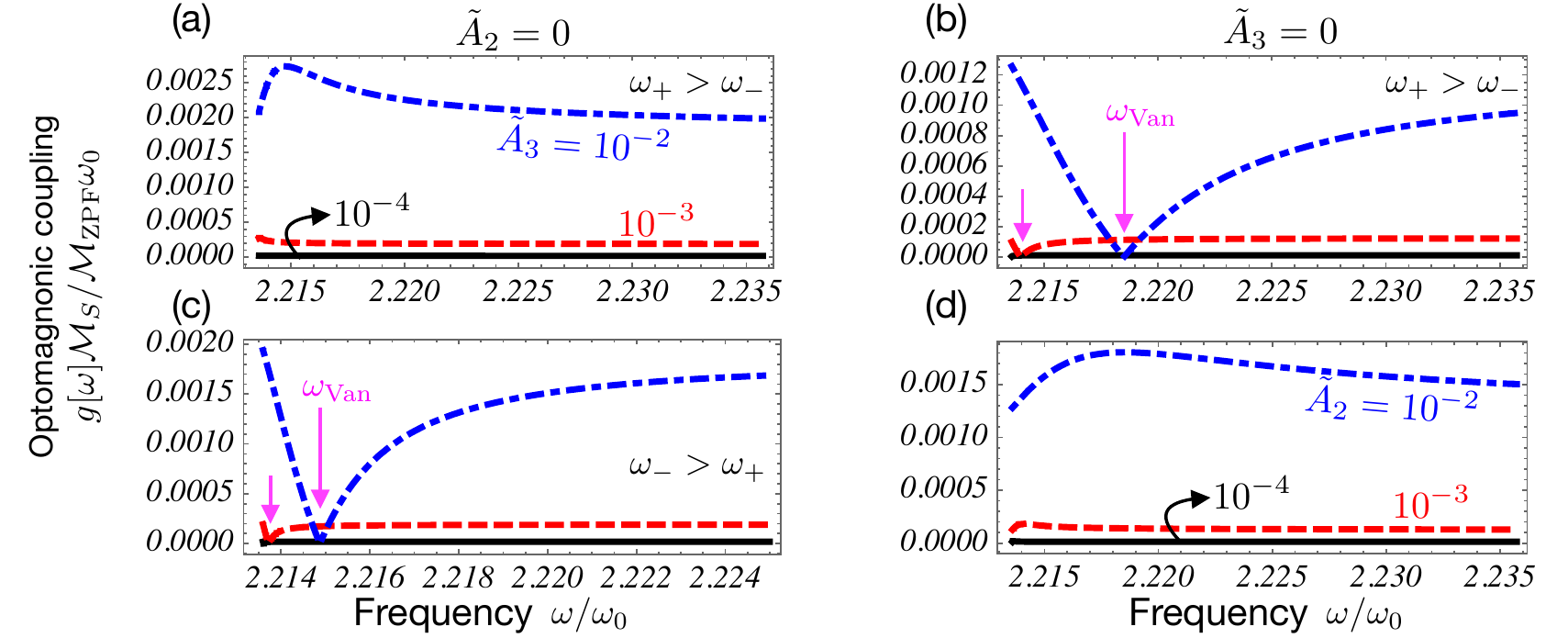}\caption{Optomagnonic coupling between non-degenerate modes as a function of the frequency (in units of $\omega_{0}$), for (a,c) $\tilde{A}_{2}=0$ and several values of $\tilde{A}_{3}$ and for (b,d) $\tilde{A}_{3}=0$ and several values of $\tilde{A}_{2}$. The arrows indicate the frequencies $\omega_{\rm{Van}}$, see Eq.~\eqref{eq:VanishingOme} for which the coupling vanishes. Parameters in correspondence with Fig.~\ref{Fig02:Permittivity}. Frequency in units of the ionic transition frequency $\omega_0$,  and the coupling is given in units of $\mathcal{M}_{{\rm ZPF}}\omega_{0}/\mathcal{M}_{S}$.}
\par\end{centering}
\label{Fig:CouplingNonDegenerate}
\end{figure*}

In Figure \ref{Fig:CouplingNonDegenerate}, we show the optomagnonic coupling for non-degenerate modes. We consider both the case $\omega_{+}>\omega_{-}$ (a and b) and $\omega_{-}>\omega_{+}$ (c and d). In both plots we assumed $\omega_{c^{\prime}}=\omega_{c}+\omega_{m}$, with $\omega_{m}/\omega_{0}=10^{-4}$. For the typical value of $\omega_{m}=10\,{\rm GHz}$, which can be externally tuned via a magnetic field, this parameter regime corresponds to an ionic transition frequency $\omega_{0}$ in the hundreds of ${\rm THz}$, which, for instance, is the case of one pronounced resonance of the Iron ions of Yttrium-Iron Garnet \cite{Bittencourt_2021_ENZ_Lett,Crossley_faraday_1969,scott_absorptionspectra_1974}. In general, the case $\tilde{A}_{3}=0$ exhibits a stronger coupling that goes to zero slower than its counterpart with $\tilde{A}_{2}=0$. As in the previous case, the coupling strength is proportional to the ratio $\mathcal{M}_{{\rm ZPF}}/\mathcal{M}_{S}$, therefore smaller magnetic volumes yield better coupling rates under the assumption of perfect overlap ($\Xi = 1$ in Eqs.~\ref{eq:FullHamiltonianDegenerate}, \ref{eq:NonDegHam01} and \ref{eq:NonDegHam02}). In both cases, we see that at the frequencies labeled $\omega_{{\rm Van}}$ in Figs.~\ref{Fig:CouplingNonDegenerate}(b,c) at which the coupling vanishes. Those are given by Eq.~\eqref{FreqSelectRules}: $\frac{\mathcal{F}[\omega_{{\rm Van}}]\mathcal{M}_{S}}{\varepsilon[\omega_{{\rm Van}}]}-1=0$, for $\omega_{+}>\omega_{-}$ and $\frac{\mathcal{F}[\omega_{{\rm Van}}]\mathcal{M}_{S}}{\varepsilon[\omega_{{\rm Van}}]}+1=0$ for $\omega_{-}>\omega_{+}$.

\begin{figure*}
\begin{centering}
\includegraphics[width=2.\columnwidth]{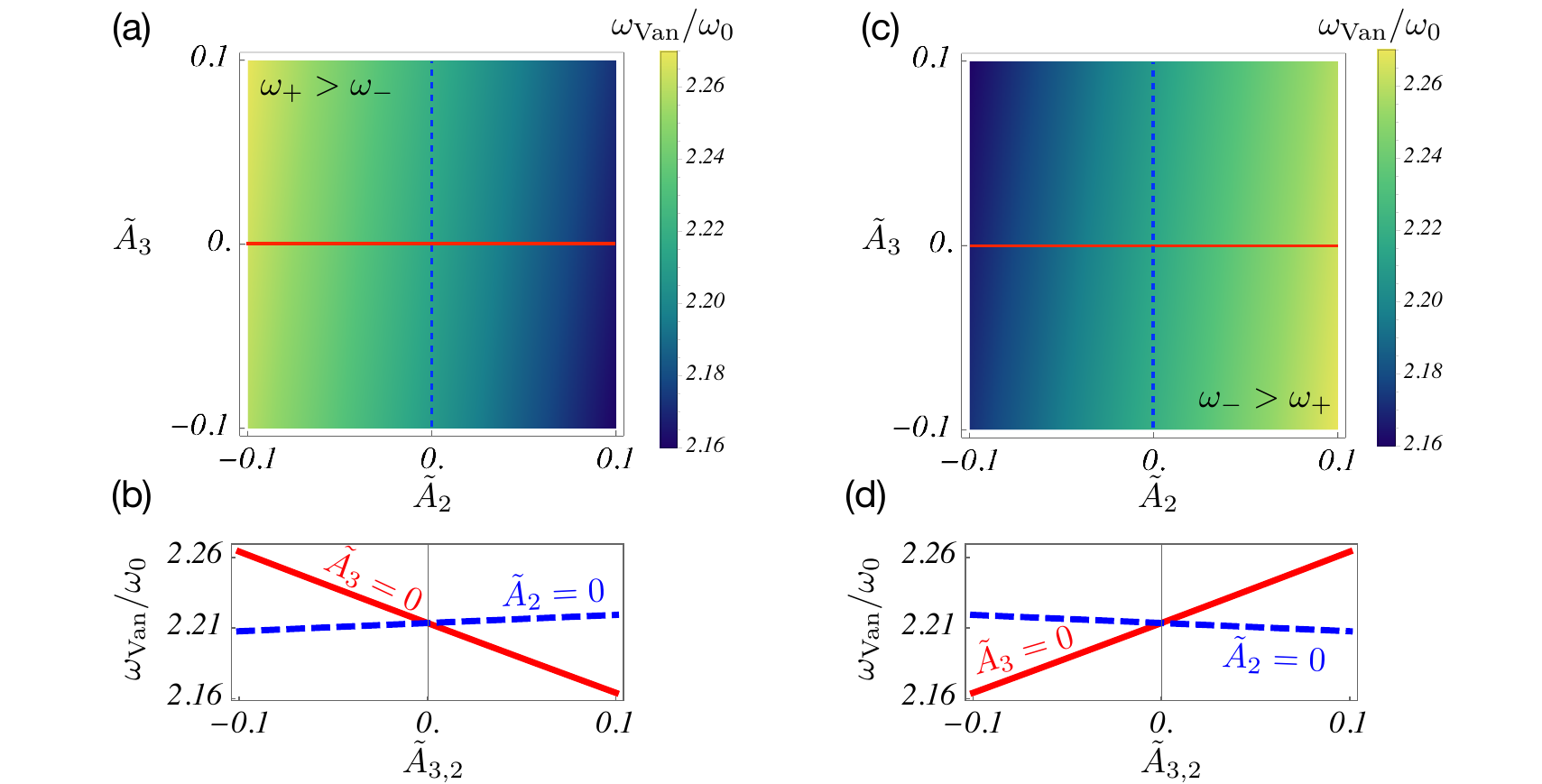}
\par\end{centering}
\begin{centering}
\caption{Frequencies at which the optomagnonic coupling between non-degenerate modes vanishes as a function of the model parameter $\tilde{A}_{2}$ and $\tilde{A}_{3}$. (a,b) for the case $\omega_{+}>\omega_{-}$ and (c,d) for the case $\omega_{-}>\omega_{+}$. Parameters in correspondence with Fig.~\ref{Fig02:Permittivity}.}
\label{Fig:VanishingCouplings}
\par\end{centering}
\end{figure*}

The frequency at which the coupling vanishes depend on the dispersion model parameters, as we show in Fig.~\ref{Fig:VanishingCouplings} we show the frequency $\omega_{{\rm Van}}$ at which the optomagnonic coupling vanishes for each case as a function of the model parameters $\tilde{A}_{3}$ and $\tilde{A}_{2}$. In these plots we restrict the parameters to the range $-0.1<\tilde{A}_{2,3}<0.1$, and we notice that in such parameter range, $\omega_{{\rm Van}}$ varies linearly with the constants $\tilde{A}_{2,3}$. Furthermore, we notice that the plots depicted in Fig.~\ref{Fig:VanishingCouplings}(a) and (c) are mirror images of each other. For the dispersion model considered here, $\omega_{{\rm Van}}$ is defined by
\begin{widetext}
\begin{equation}
\begin{aligned}
\omega_{{\rm Van}}(\omega_{0}^{2}-\omega_{{\rm Van}}^{2})\tilde{A}_{2}+\omega_{0}\tilde{A}_{3} =\pm(\omega_{0}^{2}-\omega_{{\rm Van}}^{2})\Big[\omega_{0}^{2}(\bar{\varepsilon}-1)+(\omega_{0}^{2}-\omega_{{\rm Van}}^{2})\Big],
\end{aligned}
\label{eq:VanishingOme}
\end{equation}
\end{widetext}
and thus the equation for the $-$ sign can be obtained by the transformation $\tilde{A}_{2,3}\rightarrow-\tilde{A}_{2,3}$ from the equation for the $+$ sign. This corresponds to a mirror transformation in the $(\tilde{A}_{2},\tilde{A}_{3})$ plane. 

\section{Conclusions\label{sec:Conclusions}}

To summarize, we derived a general framework for describing the coupling between magnons and photons in media with a dispersive permittivity. We applied the phenomenological quantization procedure by Milonni \cite{Milonni_1995_Field_Quantization}, generalizing it to include the Faraday effect, thus obtaining the interaction Hamiltonian describing the coupling between a uniform magnon mode and plane-wave-like optical modes in an ENZ Faraday-active medium. Our approach is valid for quasi-monochromatic fields (e.g., those of an optical cavity), under the assumption of small losses. We derived in detail the optomagnonic coupling enhancement at the ENZ frequency shown in \cite{Bittencourt_2021_ENZ_Lett} for degenerate optical modes. For non-degenerate modes, dispersion yields frequencies at which the coupling constant vanishes due to the polarization of the optical modes at those specific frequencies, giving rise to dispersion-sensitive selection rules. We discussed moreover the propagation regimes of plane-waves in dispersive Faraday active dielectric media, and showed that in both Voigt and Faraday configurations, there is a frequency range in which optical isolation can be realized via dispersion. 
Furthermore, the wave-vector of the longitudinal  plane waves in the Voigt mode diverge, a behavior that is not imprinted in the coupling constant.

Even though our results are valid for either bulk propagation or a Fabry-P\'{e}rot cavity, the formalism can be generalized to different geometries. An open cavity can be treated by incorporating quasi-normal modes \cite{Franke_2019_Quantization_of_Quasinormal,Franke_2020_quantizedquasinormal}. In this case, dispersion due to cavity design can be relevant, and the frequencies where the system exhibit a ENZ behavior can be tailored.

Our results are general and can be applied to different media, including magnetized plasmas \cite{Shen_analogof_2019} and structured media, such as gryoelectric-dielectric heterostructures \cite{Tsakmakidis_2017_breaking_lorentz}. An ENZ-based platform for magnon-photon coupling as proposed in our work combines the singular phenomena typical of waves in ENZ media, such as field concentration, energy tunneling, and wavelength stretching, with the unique characteristics of magnonic systems, for example high tunability and non-reciprocity. Applications of the platform can take into advantage the strong magnon-photon coupling for, for instance, quantum non-demolition measurements of magnons.

\textit{Acknowledgements} - The authors thank R. Boyd and O. Reshef for useful discussions. V.A.S.V. Bittencourt and S. Viola Kusminskiy acknowledge financial support from the Max Planck Society. I.L. acknowledges support from ERC Starting Grant 948504, Ram\'{o}n y Cajal fellowship RYC2018-024123-I and project RTI2018-093714-301J-I00 sponsored by MCIU/AEI/FEDER/UE.

\bibliographystyle{apsrev4-1}
\bibliography{Bibliography_Microscopic}

\end{document}